\documentclass[epj]{svjour}
\usepackage{graphics}
\usepackage{color}

\newcommand{\doubletilde}{\check}
\def\gtrapprox{\raise2.5pt\hbox{$>$}\llap{\lower2.5pt\hbox{$\sim$}}}

\begin{document}
\title{Flow curves of colloidal dispersions close to the glass transition}
\subtitle{Asymptotic scaling laws in a schematic model of mode
coupling theory}
\author{David Hajnal\inst{1} \and Matthias Fuchs\inst{2}% etc
}                     % Do not remove
\institute{Institut f\"ur Physik, Johannes Gutenberg-Universit\"at Mainz, Staudinger Weg 7,
 D-55099 Mainz, Germany \and Fachbereich Physik, Universit\"at Konstanz,
 D-78457 Konstanz, Germany}
\date{Received: date / Revised version: date}
\abstract{The flow curves, viz.~the curves of stationary stress under steady shearing, are obtained close to the glass transition in
dense colloidal dispersions using asymptotic expansions in the
schematic $F_{12}^{(\dot\gamma)}$-model of mode coupling theory.
The shear thinning of the  viscosity in fluid states and  the
yielding of  glassy states is discussed. At the transition between fluid and shear-molten glass, simple and
generalized Herschel-Bulkley laws are derived with power law
exponents that can be computed for different particle interactions
from the equilibrium structure factor.
 % end of PACS codes
} %end of abstract
\maketitle

\section{Introduction}

The behavior of dense colloidal dispersions under flow is especially
interesting at high concentrations, where flow interferes with
solidification \cite{russel,larson}. The curves of shear stress
versus shear rate under steady  flow, also called flow curves,
provide insights into the cooperative particle rearrangements under
strong external drive, and may be affected by yielding, shear
thinning and thickening, and flow instabilities leading to
heterogeneous and/or intermittent flow
\cite{Bes:07,Var:04,Var:06,Ben:96,Gan:06,Cra:07}. While
the interplay between flow and freezing into crystalline ordered
states is of interest in dispersions of rather monodisperse
particles, the interplay of flow and arrest into metastable and
amorphous solids (the glass transition) is of relevance in more
complex dispersions, especially ones consisting of polydisperse
particles, where  (frozen in) disorder frustrates crystal packing
\cite{Pet:02b,Cra:06}. Flow curves thus can provide insights
into strongly driven fluids and glasses, because the shear rate can
easily be made much larger then the internal relaxation time
$\tau_0$, and thus flow strongly affects the structural dynamics,
viz.~the strongly cooperative particle rearrangements intrinsic to
the glass transition. Various phenomenological formulae have been
proposed to describe stationary flow curves, like the Cox-Merz rule,
the Herschel-Bulkley law, and the concept of a power-law fluid
\cite{larson}. Discussion of these phenomenological laws starting
from a microscopic description of a colloidal dispersion is aim of
the present contribution.

The indicated connection between the glassy dynamics and the
nonlinear rheology of dense dispersions raises the question for a
unified description of both phenomena; see \cite{Cra:07} and
references therein. A microscopic theoretical approach for the
shear-thinning of concentrated suspensions and the yielding of
colloidal glasses was presented \cite{Fuc:02,Fuc:05c}, which
builds on the mode coupling theory (MCT)  of idealized glass
transitions developed by G\"otze and coworkers \cite{Goe:91,Goe:92}.
This description of the glass transition in a quiescent colloidal
dispersion explains many findings observed in  especially dynamic
light scattering experiments on model colloidal dispersions
\cite{Megen93,Megen93b,Megen94,Beck99,Bartsch02,Eckert03,Goe:99},
and also gives results for the linear response viscoelastic
properties, viz.~linear  (frequency dependent) shear moduli. The
extension to strongly driven stationary states
\cite{Fuc:02}, the so-called ITT (integration through
transients) approach,  and its consecutive generalization to
arbitrary time-dependent states far from equilibrium \cite{Bra:07}
and to arbitrary flow geometries \cite{Bra:08}, has yielded a
non-Newtonian constitutive equation applicable to concentrated
dispersions in arbitrary homogeneous flows, albeit under the
approximation that hydrodynamic interactions are neglected. Within
the theory, this approximation becomes valid close to the glass
transition and for weak but nonlinear flows, where the slow
structural relaxation dominates the system properties, and where
hydrodynamic interactions only affect the over-all time scale. The
structural dynamics under flow results from a competition between
local particle hindrance (termed cage effect) and the compression /
stretching (i.e. advection) of the wavelength of fluctuations
induced by the affine particle motion with the flow.

Close to the glass transition, where the structural relaxation has
become much slower than any other intrinsic relaxation process, ITT
makes universal predictions  that may be captured in simple, so
called schematic models, where a few parameters mimic the
microscopic input into the ITT approach \cite{Fuc:03}. These
so-called vertices mimic the equilibrium structure factor which
would be required to solve the complete ITT equations. Some of the
universal predictions, like the existence of power laws and their
exponents, survive the simplifications leading to schematic models
and agree  quantitatively with the microscopic ITT results for
e.g.~hard spheres. Results from one of the simplest schematic
models, the $F_{12}^{(\dot\gamma)}$-model, have been compared to
flow curves measured in dense dispersions of colloidal hard spheres
\cite{Fuc:04,Cra:06,Cra:07,Car:08,monterey} and in simulations of
hard spheres and supercooled binary Lenard-Jones mixtures
\cite{Fuc:03b,Var:06,Hen:05}. The model explained salient points of the flow
behavior of glassy systems, like a finite dynamic yield stress
beyond the glass point. It vanishes discontinuously when going below
the glass transition, where a first Newtonian plateau appears in the
viscosity,  which is followed by strong shear thinning.

Asymptotic expansions capture the leading universal results of the
schematic model and allow to discuss the complete scenario of the
nonlinear rheology as function of shear rate and separation to the
glass transition. They thereby provide insights into the connections
between various rheological features as described by ITT. The
phenomenological laws of flow curves can be discussed. In this paper
we present the asymptotic solution of the
F$_{12}^{(\dot\gamma)}$-model, thereby supplementing its use when
fitting experimental data. The model is summarized in Sect. 2. The
nonlinear stability analysis, determining the existence of power
laws and their exponents is presented in Sect. 3. It is this
stability equation which holds in the microscopic ITT equations as
well, and which provides the link to calculate e.g.~the exponents
appearing in flow curves from first principles within ITT. Sect. 4
discusses the yield process, where the structural relaxation is
driven by the external shearing. Sects. 5 and 6 discuss the results
for the flow curves, which are summarized and brought into
comparison with experiments in Sects. 7 and  8.

\section{Schematic model}

\subsection{The transient density correlator}

The schematic $F_{12}^{\left(\dot{\gamma}\right)}$-model considers
a single typical density correlator $\Phi\left(t\right)$ obeying
a Zwanzig-Mori equation \cite{Fuc:03}:
\begin{equation}
\frac{1}{\Gamma}\dot{\Phi}\left(t\right)+\Phi\left(t\right)
+\int_{0}^{t}m\left(t-t'\right)\dot{\Phi}\left(t'\right)dt'
=0.\label{eq:zwanzig_mori}
\end{equation}
$\Gamma^{-1}$ defines the (bare) time scale of the system.
A vanishing memory kernel would lead to an exponential decay of the
correlator. The equation of motion is closed by the self consistent
approximation
\begin{equation}
m\left(t\right)=\frac{v_{1}\Phi\left(t\right)+v_{2}\Phi^{2}\left(t\right)}
{1+\left(\dot{\gamma}t/\gamma_{c}\right)^{2}}
\label{eq:memory_kernel}
\end{equation}
and the normalization $\Phi\left(t\rightarrow0\right)=1-\Gamma t$.
Increasing the particle caging is modeled by increasing the coupling
vertices $v_{i}\geq0$. The parameter $\dot{\gamma}$ is the shear
rate which leads to a strain dependent decay of the memory kernel.
The parameter $\gamma_{c}$ is a scale for the magnitude of strain
that is required in order for the accumulated strain $\dot{\gamma}t$
to matter \cite{Cra:07}.

For a vanishing shear rate, $\dot{\gamma}=0$, this model has been
suggested by G\"otze \cite{Goe:84} and describes the development of slow structural
relaxation upon increasing the coupling vertices $v_{i}$. In this case the
model shows two transition lines. For sufficiently small $v_{i}$
the nonergodicity parameter $f$ defined by the long time limit of
$\Phi\left(t\right)$ is zero. By crossing one of the transition lines,
$f$ becomes positive: $f=\Phi(t\to\infty)>0$. It is proved \cite{Goe:95} that for a vanishing shear
rate $\Phi\left(t\right)$ is completely monotone and $f$ is defined
by the maximum real solution of $f+g\left(f-1\right)=0$ with $g=m(t\to\infty)=v_{1}f+v_{2}f^{2}$.
This determines the transition lines. Vertices lying on a transition
line and the corresponding nonergodicity parameters are denoted by
$v_{i}^{c}$ and $f_{c}$. The type A transition line with $f_{c}=0$
is given by $v_{1}^{c}=1$, $0\leq v_{2}^{c}\leq1$. By crossing
this line $f$ varies continuously. The type B transition line, where
$f$ jumps from zero to a finite value $f_{c}=1-\lambda$, can be
parametrized by
$\left(v_{1}^{c},v_{2}^{c}\right)=\left(\left(2\lambda-1\right),1\right)/\lambda^{2}$
with the exponent parameter $1/2\leq\lambda<1$. At $\left(v_{1}^{c},v_{2}^{c}\right)
=\left(1,1\right)$ the type A transition line merges continuously into
the type B transition line.

In the following we only consider vertices $v_{i}=v_{i}^{c}+\delta v_{i}$
close to the type B transition line. Then the dimensionless separation
parameter measuring the distance from the transition point is defined
by
\begin{equation}
\varepsilon=\frac{f_{c}\delta v_{1}+f_{c}^{2}\delta v_{2}}{1-f_{c}}.
\label{eq:separation_parameter}
\end{equation}
With these definitions the model describes the dynamics of an ideal
glass forming system. A negative value for $\varepsilon$ defines
a liquid like state. For $\varepsilon\rightarrow0^{-}$ an increasingly
stretched plateau close to $f_{c}$ develops and the time scale
for the final decay to zero depends sensitively on $\varepsilon$. A positive
separation parameter leads to a glassy state where the nonergodicity
parameter is determined to leading order by
\begin{equation}
f\left(\varepsilon\rightarrow0^{+},\dot{\gamma}=0\right)
=f_{c}+\left(1-f_{c}\right)^{2}\sqrt{\frac{\varepsilon}{1-\lambda}}.
\label{eq:nonergodicity_parameter}
\end{equation}
A detailed discussion of the quiescent $F_{12}$-model can be found in
\cite{Goe:91}.

For non-vanishing shear rates, $\dot{\gamma}\neq0$, because of the
vanishing long time limit of the memory kernel, also $f$ vanishes
for all separation parameters. An arbitrary small but finite shear
rate melts the glass and restores ergodicity. In the liquid state
($\varepsilon<0$), small shear rates have only little effect on the
dynamics. But if the time scale for the shear induced decorrelation
becomes faster than the time scale for the structural relaxation,
then the long time dynamics is shear dominated and shows the same
scaling behavior as for the glassy state ($\varepsilon>0$).

\subsection{Flow curves}

In the framework of the schematic model the steady state shear stress
$\sigma=\dot{\gamma}\eta$, where $\eta$ denotes the viscosity, follows
via integrating up the generalized shear modulus \cite{Cra:06}:
\begin{equation}
\sigma=\dot{\gamma}\int_{0}^{\infty}g\left(t,\dot{\gamma}\right)dt,
\label{eq:shear_stress}
\end{equation}
\begin{equation}
g\left(t,\dot{\gamma}\right)=v_{\sigma}\Phi^{2}\left(t\right)+\eta_{\infty}
\delta\left(t+0^-\right).
\label{eq:generalized_modulus}
\end{equation}
The small negative quantity $0^-$ in Eq. (\ref{eq:generalized_modulus}) ensures that the
integral in Eq. (\ref{eq:shear_stress}) includes the singularity of the $\delta$-function completely.
The quantity $v_{\sigma}$ is assumed to be a constant and the
parameter $\eta_{\infty}$ models viscous processes which require no
structural relaxation \cite{Cra:07}. The resulting linear moduli,
the storage $G'(\omega)$ and the loss modulus $G''(\omega)$, in the
quiescent situation are discussed in Ref.~\cite{Cra:07}. We do not
comment further on the Cox-Merz rule for which there is no basis  in
the asymptotic expansions of the $F_{12}^{(\dot\gamma)}$-model.

We sketch some basic properties of the flow curves ($\sigma$ as a
function of $\dot{\gamma}$). In the liquid state, because of the
existence of $\eta_{0}=
\int_{0}^{\infty}g\left(t,\dot{\gamma}=0\right)dt$, the flow curves
show linear asymptotes for small shear rates and the yield stress
$\sigma^{+}$, the zero shear limit of the stress
$\sigma^+=\sigma\left(\dot{\gamma}\to0\right)$, is zero. A simple Taylor expansion of
$g\left(t,\dot{\gamma}\right)$ generalizes this statement:
$\sigma=\dot{\gamma}\left(\eta_{0}+\eta_{\infty}\right)+\mathcal{O}\left(\dot{\gamma}^{3}\right)$.
The Newtonian viscosity $\eta_0$ diverges at the glass transition
according to the well known MCT power law, $\eta_0\propto
\left|\varepsilon\right|^{-\gamma}$ \cite{Goe:91}; see Eq.
(\ref{eq:alpha_timescale}) below.
In the glassy state, the time scale for the shear induced decay
scales with $1/\left|\dot{\gamma}\right|$ which leads to a non
vanishing yield stress. At $\varepsilon=0$ the yield stress jumps
from zero to a finite value. For high shear rates, all flow curves
converge to the common linear asymptote
$\dot{\gamma}\left(v_{\sigma}/2\Gamma+\eta_{\infty}\right)$  where
$v_{\sigma}/2\Gamma<\eta_{0}$. Hence the model describes the
transition from a shear thinning liquid to a yielding glass
\cite{Fuc:03}: In the liquid state, the viscosity starts at the
plateau value $\eta_{0}+\eta_{\infty}$ for small shear rates and
decays monotonically to a second plateau value
$v_{\sigma}/2\Gamma+\eta_{\infty}$ for high shear rates. In the
glassy state, the viscosity diverges in the zero shear limit.

\subsection{Parameters}

The contribution of the parameter $\eta_{\infty}$ is only an additional constant
to the viscosity which is not of interest for our work. Hence we choose $\eta_{\infty}=0$.
In the following we choose for all numerical examples $v_{2}=v_{2}^{c}=2$
and $v_{1}=v_{1}^{c}+\varepsilon\left(1-f_{c}\right)/f_{c}$.
For $\eta_{\infty}=0$ we easily verify:
\begin{equation}
\Phi\left(\varepsilon,\dot{\gamma},\gamma_{c},\Gamma,t\right)=
\breve{\Phi}\left(\varepsilon,\frac{\dot{\gamma}}{\gamma_{c}\Gamma},
\Gamma t\right),\label{eq:scaling_phi}
\end{equation}
\begin{equation}
\sigma\left(\varepsilon,\dot{\gamma},\gamma_{c},\Gamma\right)=
v_{\sigma}\gamma_{c}\breve{\sigma}\left(\varepsilon,\frac{\dot{\gamma}}
{\gamma_{c}\Gamma}\right).\label{eq:scaling_sigma}
\end{equation}
Hence, without loss of generality, we choose units in such a way that
$\gamma_{c}=\Gamma=v_{\sigma}=1$ and that $t$, $\dot{\gamma}$ and
$\sigma$ are dimensionless.
For the numerical solution of the $F_{12}^{\left(\dot{\gamma}\right)}$-model
we use the algorithm first published in \cite{Hof:91} with a grid size of $N=256$ and an
initial step size of $h=10^{-6}$, see also \cite{Haj:07} for details.
All numerical parameters, which are
relevant for our work, are tabulated in Appendix \ref{sec:num}.

\section{Glass stability analysis}

\subsection{The equation of motion}

\begin{figure}
\resizebox{1\columnwidth}{!}{\includegraphics{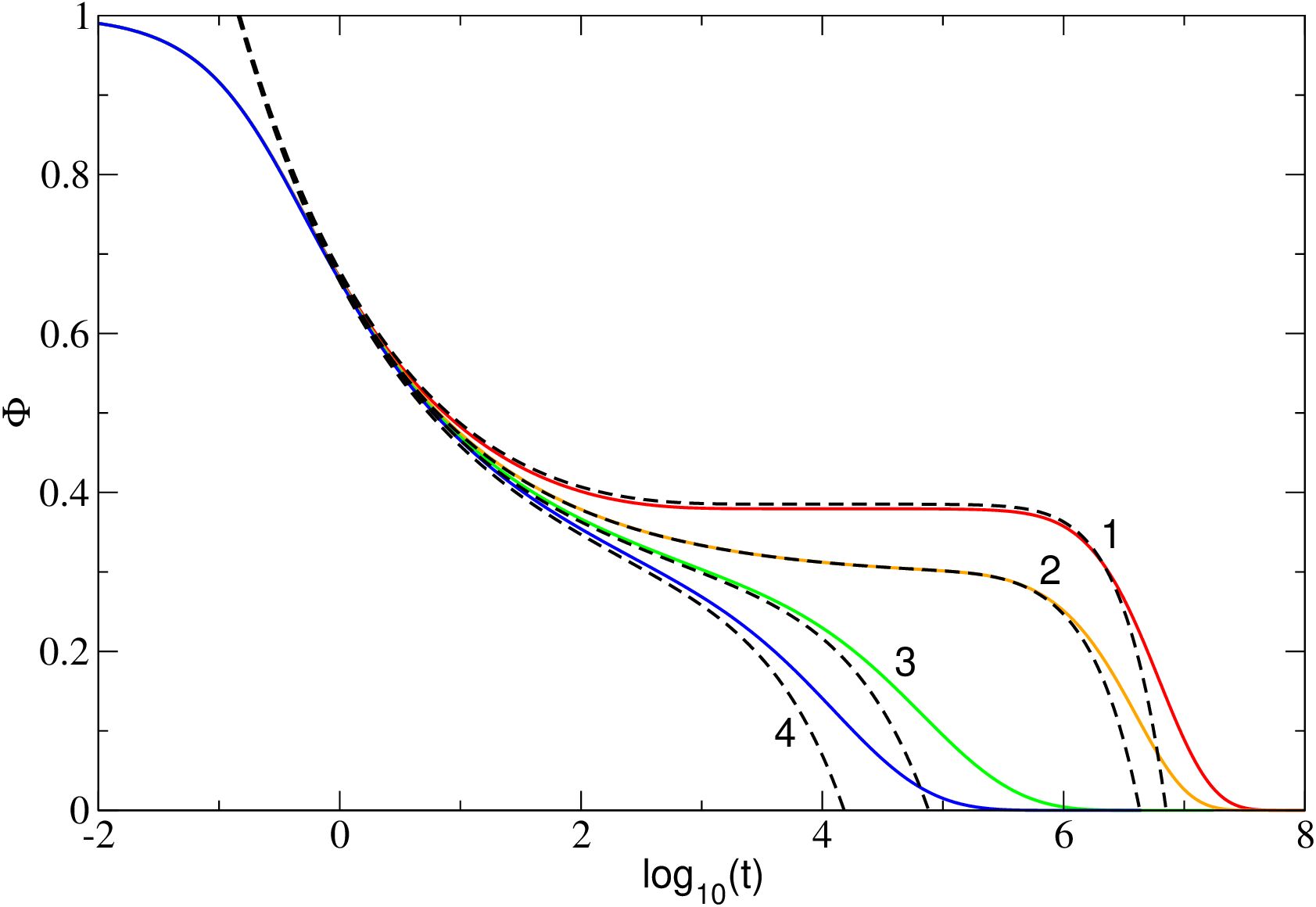}}
\caption{\label{fig:fig_1} Numerically obtained transient
correlators $\Phi\left(t\right)$ (solid lines) for $\varepsilon=0.01$ (red, 1), $\varepsilon=0$
(orange, 2), $\varepsilon=-0.005$ (green, 3) and $\varepsilon=-0.01$ (blue, 4).
All curves were calculated with $\dot{\gamma}=10^{-7}$. The dashed lines show the
corresponding numerically obtained functions $f_{c}+\left(1-f_{c}\right)^{2}\mathcal{G}\left(t\right)$.}
\end{figure}

For small separation parameters and shear rates the correlators develop
a stretched dynamics located around the critical plateau value $f_{c}$.
The discussion of the dynamics around this plateau, also called the $\beta$-relaxation process \cite{Goe:91}, is topic of this section.
To derive the equation of motion to leading order in the separation
parameter and the shear rate we first rewrite the convolution integral
in Eq. (\ref{eq:zwanzig_mori}) as $\int_{0}^{t}m\left(t-t'\right)\dot{\Phi}\left(t'\right)dt'
=\frac{d}{dt}\int_{0}^{t}m\left(t-t'\right)\Phi\left(t'\right)dt'-m\left(t\right)$.
Eq. (\ref{eq:nonergodicity_parameter}) motivates the expansion
\begin{equation}
\Phi\left(t\right)=f_{c}+\left(1-f_{c}\right)^{2}\mathcal{G}\left(t\right)+\dots
\label{eq:factorization}
\end{equation}
with $\mathcal{G}\left(t\right)=\mathcal{O}\left(|\varepsilon|^{1/2}\right)$. The function $\mathcal{G}\left(t\right)$ is often called
 $\beta$-correlator. By expanding the memory kernel in powers
of $\dot{\gamma}$ and neglecting the time derivative we obtain to
leading order the nonlinear stability equation
% which is also called the $\beta$-scaling equation
\cite{Fuc:03},
\begin{equation}
\varepsilon-c^{\left(\dot{\gamma}\right)}\left(\dot{\gamma}t\right)^{2}
+\lambda\mathcal{G}^{2}\left(t\right)=\frac{d}{dt}
\int_{0}^{t}\mathcal{G}\left(t-t'\right)\mathcal{G}\left(t'\right)dt'
\label{eq:beta_scaling}
\end{equation}
with $c^{\left(\dot{\gamma}\right)}=\left(1-\lambda\right)/\lambda^{2}$.
Eq. (\ref{eq:beta_scaling}) has to be completed with the initial
condition
\begin{equation}
\mathcal{G}\left(t\rightarrow0\right)=\left(\frac{t}{t_{0}}\right)^{-a},
\label{eq:critical_decay}
\end{equation}
where the exponent $a$ obeys
\begin{equation}
\frac{\Gamma^{2}\left(1-a\right)}{\Gamma\left(1-2a\right)}=\lambda.
\label{eq:critical_exponent}
\end{equation}
The time scale $t_{0}$ enables one to match Eq.~(\ref{eq:critical_decay}) to the microscopic dynamics in the correlators.  It has to be determined numerically by matching to the full solution for $\Phi(t)$ from Eq. (\ref{eq:zwanzig_mori}). For
$\varepsilon=0$ and $\dot{\gamma}=0$, the power law occurring in
Eq. (\ref{eq:critical_decay}) is a special solution of Eq. (\ref{eq:beta_scaling})
and describes the critical decay of $\Phi\left(t\right)$ to $f_{c}$ \cite{Goe:91}.
Fig. \ref{fig:fig_1} shows representative examples.

\subsection{The two-parameter scaling law}

One of the important aspects of $\mathcal{G}\left(t\right)$ are its homogeneity
properties \cite{Fuc:03} which follow from Eq. (\ref{eq:beta_scaling}). With some arbitrary scale $\Omega>0$ it obeys the two-parameter scaling law
\begin{equation}
\mathcal{G}\left(t,\varepsilon,\dot{\gamma}\right)=\Omega^{a}\hat{\mathcal{G}}\left(\hat{t}=
\Omega\frac{t}{t_{0}},\hat{\varepsilon}=\varepsilon\Omega^{-2a},\hat{\dot{\gamma}}=
\dot{\gamma}t_{0}\Omega^{-\left(1+a\right)}\right)
\label{eq:scaling_law}
\end{equation}
with the master function obeying the initial condition $\hat{\mathcal{G}}\left(\hat{t}\rightarrow0\right)=\left(\hat{t}\right)^{-a}$.
Eq. (\ref{eq:scaling_law}) allows to define three regions. By choosing
$\Omega=\left|\dot{\gamma}t_{0}\right|^{\frac{1}{1+a}}$ we obtain
$\left|\hat{\dot{\gamma}}\right|=1$ and $\hat{\varepsilon}=\varepsilon/\varepsilon_{\dot{\gamma}}$
with the natural scale
\begin{equation}
\varepsilon_{\dot{\gamma}}=\left|\dot{\gamma}t_{0}\right|^{\frac{2a}{1+a}}.
\label{eq:epsilon_scale}
\end{equation}
The liquid region is defined by $\varepsilon\ll-\varepsilon_{\dot{\gamma}}$,
where a finite shear rate barely distorts $\mathcal{G}\left(t\right)$. In the
transition region, $\left|\varepsilon\right|\ll\varepsilon_{\dot{\gamma}}$,
the dynamical anomalies are the most pronounced. The yielding glass
region is defined by $\varepsilon\gg\varepsilon_{\dot{\gamma}}$. These
definitions hold for $\left|\varepsilon\right|\ll1$ and $\left|\dot{\gamma}t_{0}\right|\ll1$
what we assume throughout in the following.

The present two-parameter scaling law bears some similarity to the one presented by G\"otze and Sj\"ogren for the description of thermally activated processes in   glasses \cite{Hil:92}. In both cases, ideal glass states are destroyed by additional decay mechanisms. Yet, the ITT equations and the generalised MCT equations differ qualitatively in their description of the states at $\varepsilon\ge0$. While the generalised MCT there predicts  finite Newtonian viscosities, ITT predicts  yield-stresses. Also both mechanisms can appear simultaneously and then need to be combined \cite{Cra:07}. The similarity between both two-parameter scaling laws thus underlines the universality of the glass stability analysis, which is determined by quite fundamental principles. In Eq. (\ref{eq:beta_scaling}), the shear rate can only be a relevant perturbation (at long times) if it appears multiplied by time itself. Symmetry dictates the appearance of $(\dot\gamma t)^2$, because the sign of
 the shear rate must not matter. The aspect that shear melts the glass determines the negative sign of $(\dot\gamma t)^2$.

\subsection{The power series ansatz}

\begin{figure}
\resizebox{1\columnwidth}{!}{\includegraphics{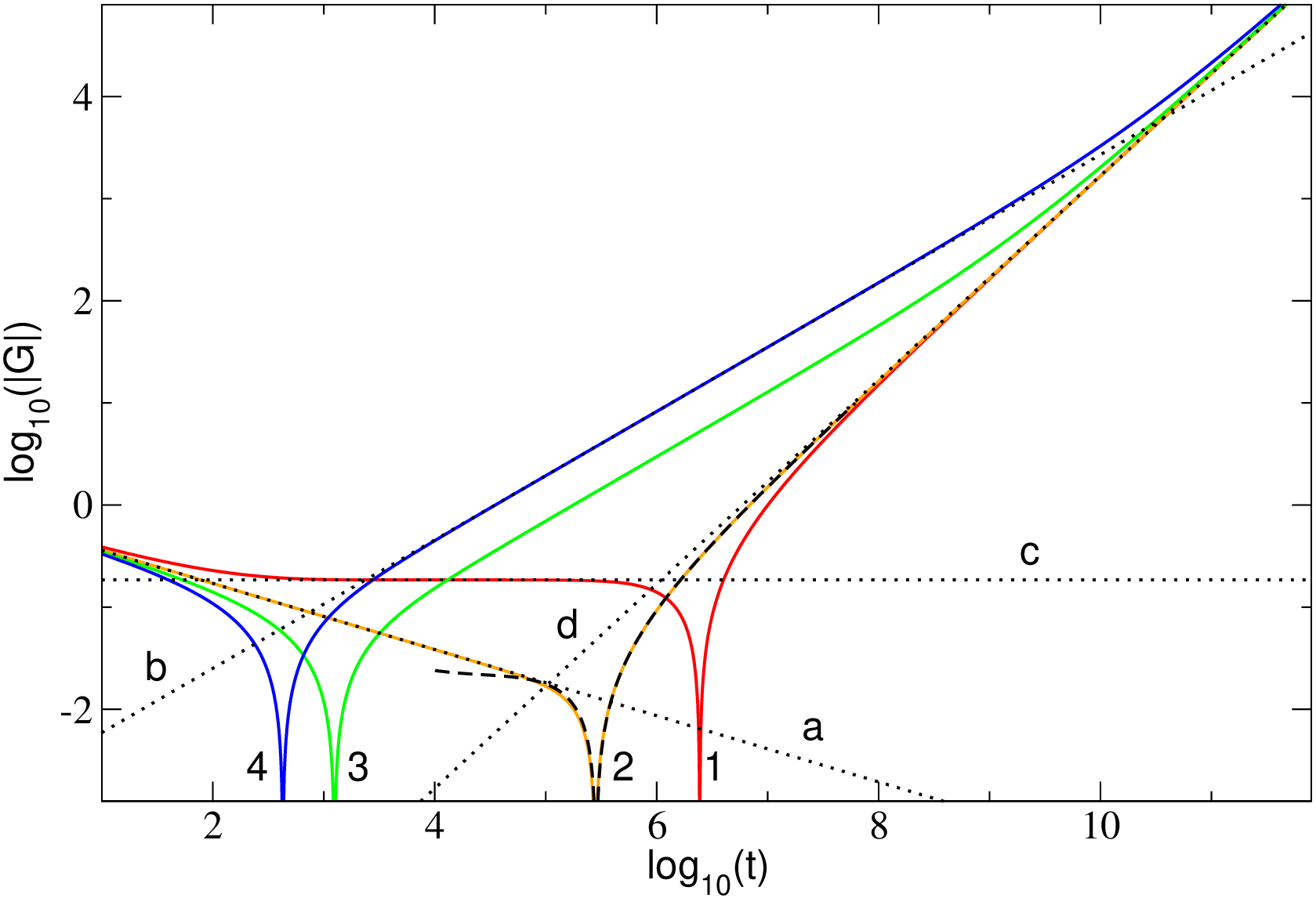}}
\caption{\label{fig:fig_2} An overview of the properties of
$\mathcal{G}\left(t\right)$ (solid lines) for the same values for $\varepsilon$ and $\dot{\gamma}$ (1-4)
as in Fig. \ref{fig:fig_1}. The dotted lines show the leading asymptotes
for the corresponding time scales: The critical decay $\left(t/t_{0}\right)^{-a}$
(a), the von Schweidler law $-\left(t/\tau_{0}\right)^{b}$ (b),
the arrest on the plateau value $\sqrt{\varepsilon/\left(1-\lambda\right)}$
(c) and the shear-induced linear asymptote $-t/\tau_{\dot{\gamma}}$ (d).
The dashed line shows Eq. (\ref{eq:long_time_expansion}) evaluated to $n=3$ for
$a_{1}=-1.39\cdot 10^{3}$ ($\varepsilon=0$).}
\end{figure}

Eq. (\ref{eq:beta_scaling}) can be solved by using a generalized
power series ansatz
\begin{equation}
\mathcal{G}\left(t\right)=\alpha t^{u}\left(1+\sum_{n}a_{n}t^{\nu n}\right),
\label{eq:power_series}
\end{equation}
where the exponents $u$, $\nu$ and the coefficients $\alpha$, $a_{1}$ obey
\begin{equation}
\varepsilon-c^{\left(\dot{\gamma}\right)}\left(\dot{\gamma}t\right)^{2}=
\alpha^{2}\Gamma_{0,0}t^{2u}+2\alpha^{2}a_{1}\Gamma_{0,1}t^{2u+\nu},
\label{eq:coefficients}
\end{equation}
with $\Gamma_{n,n'}=\frac{\Gamma\left(1+u+n\nu\right)\Gamma\left(1+u+n'\nu\right)}
{\Gamma\left(1+2u+\left(n+n'\right)\nu\right)}-\lambda$
and the other coefficients are determined by the recursion formula
$a_{n\geq2}=-\frac{1}{2\Gamma_{0,n}}\sum_{n'=1}^{n-1}\Gamma_{n-n',n'}a_{n-n'}a_{n'}$.

Along the lines of the work \cite{Hil:92},
we have analyzed all possible combinations that solve Eq. (\ref{eq:coefficients}) in detail, see \cite{Haj:07}.
Each defines an (asymptotic) series expansion describing the dynamics
in a special regime on a special time scale. There are four solutions
for the liquid region, two for the transition region and three solutions
describing yielding glassy dynamics. Here we only present a qualitative
overview containing the leading asymptotes. In the liquid region,
for short times $\mathcal{G}\left(t\right)$ follows $\left(t/t_{0}\right)^{-a}$
and merges into a second power law $-\left(t/\tau_{0}\right)^{b}$
for intermediate times with the von Schweidler exponent $b$ obeying
\begin{equation}
\frac{\Gamma^{2}\left(1+b\right)}{\Gamma\left(1+2b\right)}=\lambda.
\label{eq:von_schweidler}
\end{equation}
The two-parameter scaling law leads to
\begin{equation}
\tau_{0}=\hat{\tau}_{0}t_{0}\left|\varepsilon\right|^{-\gamma},
\label{eq:alpha_timescale}
\end{equation}
\begin{equation}
\gamma=\frac{a+b}{2ab}.
\end{equation}
The parameter $\hat{\tau}_{0}$ has to be determined numerically.
Without shear, the divergence of $\tau_0$ from Eq.
(\ref{eq:alpha_timescale}) is the origin of the divergence of the
Newtonian viscosity
\begin{equation}\label{viscosity}
\eta_0 = G_\infty^c \tau_0 \hat \tau_\eta \;,
\end{equation}
where $G_\infty^c$ denotes the shear modulus (elastic constant) of the
glass \cite{Cra:07}, and $\hat\tau_\eta$ is a constant. For long times
$\mathcal{G}\left(t\right)$ merges into the linear asymptote
$-t/\tau_{\dot{\gamma}}$ with
\begin{equation}
\tau_{\dot{\gamma}}=\frac{1}{\left|\dot{\gamma}\right|}\sqrt{\frac{\lambda-\frac{1}{2}}
{c^{\left(\dot{\gamma}\right)}}}.
\label{eq:tau_gammadot}
\end{equation}
In the transition region, after following $\left(t/t_{0}\right)^{-a}$ the function
$\mathcal{G}\left(t\right)$ merges into the long time asymptote $-t/\tau_{\dot{\gamma}}$.
In the yielding glass region, $\mathcal{G}\left(t\right)$ follows $\left(t/t_{0}\right)^{-a}$,
arrests on the plateau value $\sqrt{\varepsilon/\left(1-\lambda\right)}$
for intermediate times and merges into the linear asymptote $-t/\tau_{\dot{\gamma}}$
for long times. So we can summarize that the short- and long time
asymptotes are common for all $\varepsilon$ if $\dot{\gamma}\neq0$
is common.
Fig. \ref{fig:fig_2} shows an overview of the properties of $\mathcal{G}\left(t\right)$.

\subsection{The long time dynamics}

We briefly present the special asymptotic series expansion which is essential
to understand the asymptotic behavior of the flow curves. For long
times, $\varepsilon$ can be neglected on the left side of Eq. (\ref{eq:coefficients}).
We solve the resulting equation by choosing $u=1$ and $\Gamma_{0,1}=0$.
The result can be written as
\begin{equation}
\mathcal{G}\left(t\right)=-\frac{t}{\tau_{\dot{\gamma}}}\left(1+\sum_{n}a_{n}^{*}\left(a_{1}t^{-c}\right)^{n}\right)
\label{eq:long_time_expansion}
\end{equation}
with $a_{1}^{*}=1$, $a_{n\geq2}^{*}=-\frac{1}{2\Gamma_{0,n}}\sum_{n'=1}^{n-1}\Gamma_{n-n',n'}a_{n-n'}^{*}a_{n'}^{*}$
and the exponent $c$ is defined by
\begin{equation}
c=\frac{2\lambda-1}{\lambda}.
\label{eq:c_exponent}
\end{equation}
A numerical example for $\varepsilon=0$ is shown in Fig. \ref{fig:fig_2}.

The coefficient $a_{1}$ has to be determined by matching to shorter
time scales \cite{Fuc:03}. It obeys the two-parameter scaling law
\begin{equation}
a_{1}\left(\varepsilon,\dot{\gamma}\right)= \hat{a}_{1}\left(\varepsilon\Omega^{-2a},
\dot{\gamma}t_{0}\Omega^{-\left(1+a\right)}\right)\left(\frac{\Omega}{t_{0}}\right)^{-c}.
\label{eq:a1_coefficient}
\end{equation}
Because of this scaling law, the non-trivial factor of ${a}_{1}$ depends on the ratio $\varepsilon/\varepsilon_{\dot{\gamma}}$ only:
\begin{equation}
a_{1}\left(\varepsilon,\dot{\gamma}\right)=\left(t_{0}\right)^{c}
\left|\dot{\gamma}t_{0}\right|^{-\frac{c}{1+a}}\hat{a}_{1}\left(\varepsilon/\varepsilon_{\dot{\gamma}},1\right).
\label{eq:a1_coefficient_2}
\end{equation}
This behavior will be crucial in the following
as it enables us to describe the flow curves for small $\left|\varepsilon\right|$ and
$\left|\dot{\gamma}t_{0}\right|$. Three limiting behaviors of ${a}_{1}$ can be determined, by identifying the time scales at shorter times, to which the series from Eq. (\ref{eq:long_time_expansion}) needs to be matched.

\subsubsection{Liquid}

In the liquid region, the crossover from the von Schweidler dynamics to the shear-induced long time dynamics
is of importance. To obtain the appropriate series expansion describing this crossover, we
solve Eq. (\ref{eq:coefficients}) by neglecting $\varepsilon$ and
choosing $\alpha=-\left(\tau_{0}\right)^{-b}$ and $u=b$. The resulting natural time scale
for the crossover process is given by
$\tau_{b}=\tau_{0}\left|\dot{\gamma}\tau_{0}\right|^{-\frac{1}{1-b}}$
to which the long time series has to be matched. For this, we choose $\Omega$ such that
$\hat{t}=t/\tau_{b}$, use Eq. (\ref{eq:alpha_timescale}) and assume that in the limit
$\varepsilon/\varepsilon_{\dot{\gamma}}\rightarrow-\infty$
the coefficient $\hat{a}_{1}$ converges toward some constant. We
obtain
\begin{equation}
a_{1}\left(\varepsilon/\varepsilon_{\dot{\gamma}}\rightarrow-\infty\right)
=a_{1}^{-}\left(t_{0}\right)^{c}\left|\varepsilon\right|^{\gamma c\frac{b}{1-b}}\left|\dot{\gamma}t_{0}\right|^{-\frac{c}{1-b}}
\label{eq:a1_liquid}
\end{equation}
where $a_{1}^{-}$ is an appropriately chosen constant.

\subsubsection{Critical point}

At the glass transition point, Eq. (\ref{eq:a1_coefficient_2}) leads directly to
\begin{equation}
a_{1}\left(0,\dot{\gamma}\right)={a}_{1}^{0}\left(t_{0}\right)^{c}\left|\dot{\gamma}t_{0}\right|^{-\frac{c}{1+a}}
\label{eq:a1_transition}
\end{equation}
with some constant $a_{1}^{0}$.
A matching to the natural time scale, where the crossover from the critical short time dynamics
to the shear-induced long time dynamics occurs, leads to the same result.

\subsubsection{Yielding glass}

In the yielding glass region, the initial decay from the plateau $\sqrt{\varepsilon/\left(1-\lambda\right)}$
is relevant. To describe this,
Eq. (\ref{eq:coefficients}) can be solved by choosing $u=0$
and $\nu=2$. The resulting natural time scale for the initial decay is given
by $t_{b}=\sqrt{\varepsilon}/\left|\dot{\gamma}\right|$. The long time series has to be matched to this time scale.
With the same argument as for the liquid region, this leads to
\begin{equation}
a_{1}\left(\varepsilon/\varepsilon_{\dot{\gamma}}\rightarrow\infty\right)=
a_{1}^{+}\left(t_{0}\right)^{c}\left|\varepsilon\right|^{\frac{c}{2}}\left|\dot{\gamma}t_{0}\right|^{-c}
\label{eq:a_1_glass}
\end{equation}
where $a_{1}^{+}$ is another constant.

\section{The yield process}

\begin{figure}
\resizebox{1\columnwidth}{!}{\includegraphics{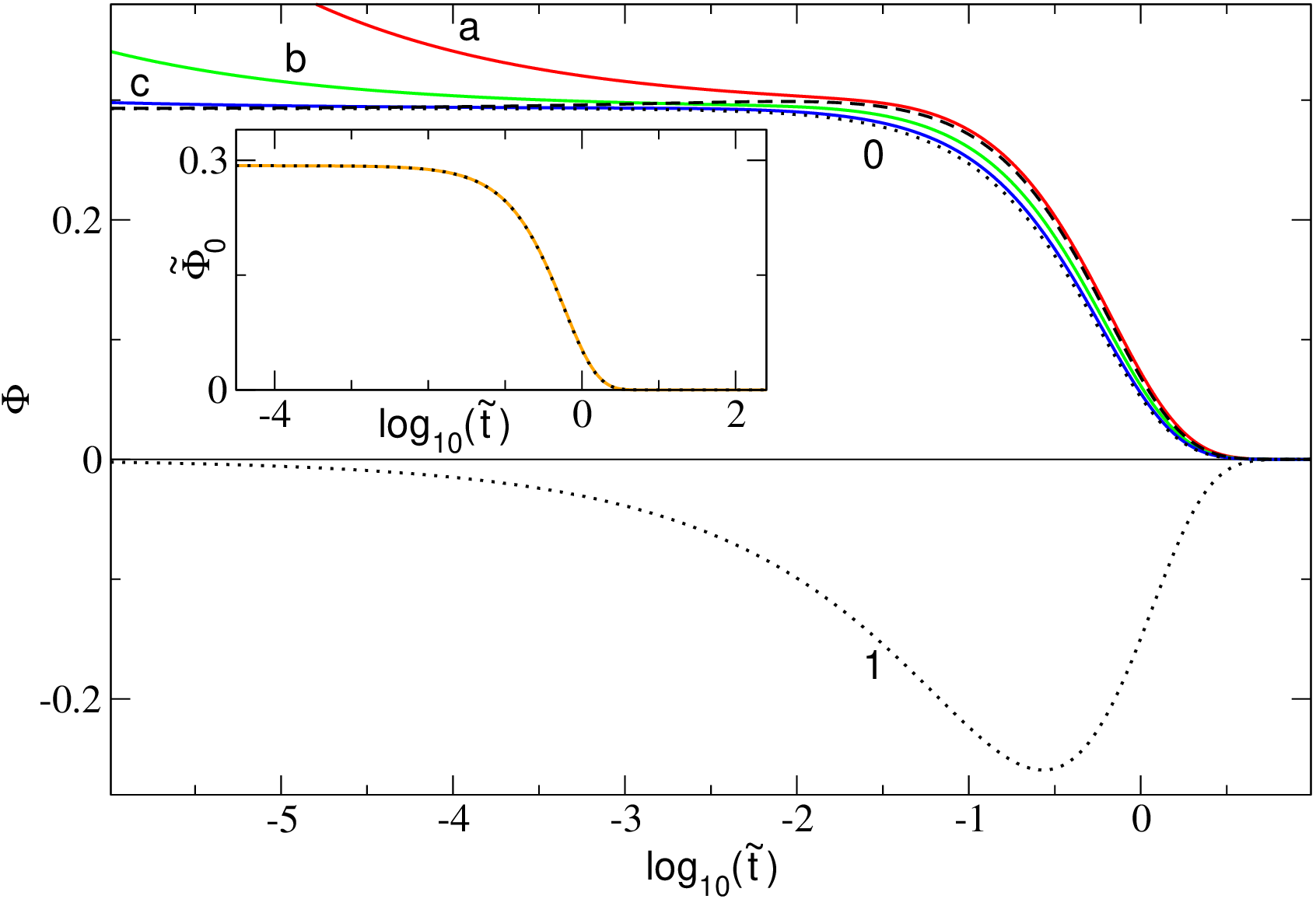}}
\caption{\label{fig:fig_a} Numerically determined master functions of the 'time-shear-superposition principle' (dotted lines) $\tilde{\Phi}_{0}\left(\tilde{t}\right)$ (0) and $\tilde{\Phi}_{1}\left(\tilde{t}\right)$ (1). The solid lines show numerically obtained transient density correlators for $\varepsilon=0$ and $\dot{\gamma}=10^{-7}$ (red, a), $\dot{\gamma}=10^{-9}$ (green, b) and $\dot{\gamma}=10^{-12}$ (blue, c), plotted as functions of the
rescaled time $\tilde{t}$. The plots demonstrate that the rescaled correlators converge to $\tilde{\Phi}_{0}\left(\tilde{t}\right)$ for $\dot{\gamma}\rightarrow 0$, the blue curve (c) is already quite close to the master curve (0). The dashed line shows $\tilde{\Phi}_{0}\left(\tilde{t}\right)+a_1\left|\dot{\gamma}\right|^{c}\tilde{\Phi}_{1}\left(\tilde{t}\right)$ for $\dot{\gamma}=10^{-7}$ and the same numerical value for $a_1$ as in Fig. \ref{fig:fig_2}. This first order expansion already describes quite well the shear-induced decay of the red curve (a). The inset demonstrates that the master function $\tilde{\Phi}_{0}\left(\tilde{t}\right)$ (dotted line) can be well approximated by the exponential function given by Eq. (\ref{eq:alpha0_exp}) (solid line). The curves overlap completely.
}
\end{figure}

While in the previous section, the dynamics of the correlator around the plateau $f_c$ was studied, now  its subsequent decay is studied. This dynamics is  called the $\alpha$-relaxation process in fluids \cite{Goe:91}, or the 'yield' process in glasses under shear.
In this section we derive equations of motion for the shear induced
final decay (yielding) of $\Phi\left(t\right)$ to zero. First we rewrite Eq.
(\ref{eq:long_time_expansion}) by introducing a natural time scale:
\begin{equation}
\tilde{\mathcal{G}}\left(\tilde{t}\right)= -\tilde{t}-\sum_{n\geq1}a_{n}^{*}\left(a_{1}
\left|\dot{\gamma}\right|^{c}\right)^{n}\left(\left|\dot{\gamma}\right|\tau_{\dot{\gamma}}
\right)^{-cn}\left(\tilde{t}\right)^{-cn+1},
\label{eq:rescaled_series}
\end{equation}
\begin{equation}
\tilde{t}=\frac{t}{\tau_{\dot{\gamma}}}.
\label{eq:rescaled_time}
\end{equation}

Now we consider Eq. (\ref{eq:zwanzig_mori}). By rescaling time and eliminating the
short time dynamics we obtain:
\begin{equation}
\tilde{\Phi}\left(\tilde{t}\right)  -  \tilde{m}\left(\tilde{t}\right)+\frac{d}{d\tilde{t}}\int_{0}^{\tilde{t}}\tilde{m}\left(\tilde{t}-\tilde{t}'\right)\tilde{\Phi}\left(\tilde{t'}\right)d\tilde{t}'=0,
\label{eq:rescaled_mori}
\end{equation}
\begin{equation}
\tilde{m}\left(\tilde{t}\right)  =  \frac{v_{1}\tilde{\Phi}\left(\tilde{t}\right)+v_{2}{\tilde{\Phi}}^{2}\left(\tilde{t}\right)}{1+\left(\tilde{\dot{\gamma}}\tilde{t}\right)^{2}},
\end{equation}
\begin{equation}
\tilde{\dot{\gamma}}  =  \left|\dot{\gamma}\right|\tau_{\dot{\gamma}}=
\sqrt{\frac{\lambda-\frac{1}{2}}{c^{\left(\dot{\gamma}\right)}}}.
\end{equation}
Note that $\tilde{\dot{\gamma}}$ does not dependent on $\dot{\gamma}$.
Because $\tilde{\dot{\gamma}}$ is a constant of the order one, we
can neglect the influence of $\varepsilon$:
\begin{equation}
\left(v_{1},v_{2}\right)  =  \left(v_{1}^{c},v_{2}^{c}\right).
\end{equation}
With this, the explicit dependence on $\varepsilon$ and $\dot{\gamma}$
is eliminated. Eq. (\ref{eq:rescaled_series}) motivates the ansatz of a generalized 'time-shear-superposition principle':
\begin{equation}
\tilde{\Phi}\left(\varepsilon,\dot{\gamma},\tilde{t}\right)  =  \sum_{n}\left[\tilde{\chi}\left(\varepsilon,\dot{\gamma}\right)\right]^{n}\tilde{\Phi}_{n}\left(\tilde{t}\right).
\label{eq:alpha_series}
\end{equation}
By substituting this in Eq. (\ref{eq:rescaled_mori}) we obtain a recursively defined
sequence of integro-differential equations:
\begin{equation}
\tilde{\Phi}_{n}\left(\tilde{t}\right)-\tilde{m}_{n}\left(\tilde{t}\right)+\frac{d}{d\tilde{t}}\int_{0}^{\tilde{t}}\sum_{n'=0}^{n}\tilde{m}_{n-n'}\left(\tilde{t}-\tilde{t}'\right)\tilde{\Phi}_{n'}\left(\tilde{t'}\right)d\tilde{t}'  =  0,
\end{equation}
\begin{equation}
\tilde{m}_{n}\left(\tilde{t}\right)  =  \frac{v_{1}^{c}\tilde{\Phi}_{n}\left(\tilde{t}\right)+v_{2}^{c}\tilde{\Psi}_{n}\left(\tilde{t}\right)}{1+\left(\tilde{\dot{\gamma}}\tilde{t}\right)^{2}},
\end{equation}
\begin{equation}
\tilde{\Psi}_{n}\left(\tilde{t}\right)  =  \sum_{n'=0}^{n}\tilde{\Phi}_{n-n'}\left(\tilde{t}\right)\tilde{\Phi}_{n'}\left(\tilde{t}\right).
\end{equation}

For short rescaled times we require $\tilde{\Phi}\left(\tilde{t}\rightarrow0\right)=f_{c}+\left(1-f_{c}\right)^{2}\tilde{\mathcal{G}}\left(\tilde{t}\right)$
with $\tilde{\mathcal{G}}\left(\tilde{t}\right)$ given by Eq. (\ref{eq:rescaled_series}).
Hence we postulate:
\begin{equation}
\tilde{\chi}\left(\varepsilon,\dot{\gamma}\right)  =  a_{1}\left(\varepsilon,\dot{\gamma}\right)\left|\dot{\gamma}\right|^{c},
\label{eq:chi}
\end{equation}
\begin{equation}
\tilde{\Phi}_{0}\left(\tilde{t}\rightarrow0\right)=f_{c}-\left(1-f_{c}\right)^{2}\tilde{t},
\label{eq:alpha_0}
\end{equation}
\begin{equation}
\tilde{\Phi}_{n\geq1}\left(\tilde{t}\rightarrow0\right)= -\left(1-f_{c}\right)^{2}a_{n}^{*}\left(\left|\dot{\gamma}\right|\tau_{\dot{\gamma}}\right)^{-cn}\left(\tilde{t}\right)^{-cn+1}.
\label{eq:alpha_1}
\end{equation}

Now we have to show that the postulated short time asymptotes represent
solutions of the equations of motions for $\tilde{t}\rightarrow0$.
For this we use the ansatz $\tilde{\Phi}_{0}\left(\tilde{t}\to0\right)=f_{c}+\alpha_{0}\tilde{t}$,
$\tilde{\Phi}_{n\geq1}\left(\tilde{t}\to0\right)=\alpha_{n}\left(\tilde{t}\right)^{-cn+1}$
and neglect $\left(\tilde{\dot{\gamma}}\tilde{t}\right)^{2}$. First
we consider $n=0$ and neglect all terms of the order $\left(\tilde{t}\right)^{2}$.
The resulting equation is satisfied for arbitrary $\alpha_{0}$. Now
we consider $n\geq1$. We can write $\tilde{\Psi}_{n}\left(\tilde{t}\right)=2\tilde{\Phi}_{0}\left(\tilde{t}\right)\tilde{\Phi}_{n}\left(\tilde{t}\right)+\sum_{n'=1}^{n-1}\tilde{\Phi}_{n-n'}\left(\tilde{t}\right)\tilde{\Phi}_{n'}\left(\tilde{t}\right)$
where the first term is of the order $\left(\tilde{t}\right)^{-cn+1}$.
The sum vanishes for $n=1$ and is of the order $\left(\tilde{t}\right)^{-cn+2}$
for $n\geq2$, a sub-leading term. Hence we can write: $\tilde{\Psi}\left(\tilde{t}\rightarrow0\right)=2\tilde{\Phi}_{0}\left(\tilde{t}\right)\tilde{\Phi}_{n}\left(\tilde{t}\right)$.
Similarly, we can show that $\frac{d}{d\tilde{t}}\int_{0}^{\tilde{t}}\sum_{n'=1}^{n-1}\tilde{m}_{n-n'}\left(\tilde{t}-\tilde{t}'\right)\tilde{\Phi}_{n'}\left(\tilde{t'}\right)d\tilde{t}'$
is of the order $\left(\tilde{t}\right)^{-cn+2}$. We substitute these
results in the equation of motion and neglect all sub-leading terms.
Then the resulting equation is satisfied for arbitrary constants $\alpha_{n}$.
We remark that integrals like $\int_{0}^{\tilde{t}}\left(\tilde{t}'\right)^{-cn+1}d\tilde{t}'$
do not exist for $n\geq2/c$, but we can formally define objects like
$\frac{d}{d\tilde{t}}\int_{0}^{\tilde{t}}\left(\tilde{t}'\right)^{-cn+1}d\tilde{t}'=\lim_{\delta\rightarrow0}\frac{d}{d\tilde{t}}\int_{\delta}^{\tilde{t}}\dots d\tilde{t}'=\left(\tilde{t}\right)^{-cn+1}$
for arbitrary $n$. Such definitions were also implicitly used for
the series expansions for the glass stability analysis.

With this we can state: We have derived a closed set of equations for
$\tilde{\Phi}_{n}\left(\tilde{t}\right)$ which, combined with the
appropriate initial conditions, can be solved recursively. The solutions
are not dependent on $\varepsilon$ and $\dot{\gamma}$. While the form of the correlators is model-dependent, their power-law initial decays follows from the universal stability equation (\ref{eq:beta_scaling}). It will be these power-law variations that determine the exponents in the dependence of the flow curves on shear-rate. We remark
that $n=0$ reproduces the result shown in \cite{Fuc:03} and for $\varepsilon=0$ Eq. (\ref{eq:alpha_series}) reduces to
the series expansion presented in \cite{Hen:05}. Fig. \ref{fig:fig_a} shows numerical results for the master functions $\tilde{\Phi}_{0}\left(\tilde{t}\right)$ and $\tilde{\Phi}_{1}\left(\tilde{t}\right)$ which are consistent with our results for the glass stability analysis. We also observe that the master function $\tilde{\Phi}_{0}\left(\tilde{t}\right)$ can be well approximated by an exponential function:
\begin{equation}
\label{eq:alpha0_exp}
\tilde{\Phi}_{0}\left(\tilde{t}\right)=f_c\exp\left(-\tilde{t}/\tilde{t}_0\right).
\end{equation}
The time scale $\tilde{t}_0=f_c/\left(1-f_c\right)^2$ follows from Eq. (\ref{eq:alpha_0}).

For the discussion of the flow curves we are especially interested
in $n<2/c$ where for some finite $\delta>0$ the integral $\int_{0}^{\delta}\tilde{\Phi}_{n}\left(\tilde{t}\right)d\tilde{t}$
exists. If we assume that $\tilde{\Phi}_{n}\left(\tilde{t}\right)$
is bounded for all $\tilde{t}\geq\delta$ and all $0\leq n<2/c$, then
the long time limit of $\tilde{m}_{n}\left(\tilde{t}\right)$ is obviously
zero. Then by analyzing the Laplace-transforms of the equations motion
we can easily show:
\begin{equation}
\lim_{\tilde{t}\rightarrow\infty}\tilde{\Phi}_{n}\left(\tilde{t}\right)  =  0.
\end{equation}
In addition, we assume for the following that the decay to zero is
such that $\tilde{\Phi}_{n}\left(\tilde{t}\right)$ is integrable
for all $0\leq n<2/c$. Then for instance, the integral $\tilde{\dot{\gamma}} \int_{0}^{\infty}{\tilde{\Phi}_{0}}^{2}\left(\tilde{t}\right)d\tilde{t}$
defines the critical yield stress at the glass transition point.

\section{Flow curves: asymptotic regime}

\subsection{The $\Lambda$-formula}

The results of the previous sections, especially Eq. (\ref{eq:alpha_series}) and Eq. (\ref{eq:chi}),
allow us to derive an analytical expression summarizing many aspects
of the flow curves.

For sufficiently small $\varepsilon$ and $\left|\dot{\gamma}\right|$, because of the stretched dynamics, the dominant contribution
to the integral in Eq. (\ref{eq:shear_stress}) is determined by the final decay of $\Phi^2\left(t\right)$ from the plateau to zero.
Eq. (\ref{eq:alpha_series}) and Eq. (\ref{eq:chi}) represent an asymptotic expansion for the shear-induced final decay of $\Phi\left(t\right)$. As demonstrated in Fig. \ref{fig:fig_a}, the first few low order terms are sufficient to describe the long
time asymptote, provided that separation parameter and shear rate are sufficiently small.
We notice that because of Eq. (\ref{eq:alpha_0})
and Eq. (\ref{eq:alpha_1}) the integrals
\begin{equation}
\sigma_{n}^{*}= \dot{\gamma}\tau_{\dot{\gamma}}\int_{0}^{\infty}\sum_{n'=0}^{n}
\tilde{\Phi}_{n-n'}\left(\tilde{t}\right)\tilde{\Phi}_{n'}\left(\tilde{t}\right)d\tilde{t}
\label{eq:integrals}
\end{equation}
exist for $n\leq\left\lfloor 2/c\right\rfloor _{-}=3$ (here
$\left\lfloor z\right\rfloor _{-}$ denotes the maximum integer number
$n$ with $n<z$). Because of the rescaling of time in Eq. (\ref{eq:rescaled_time}),
the $\sigma_{n}^{*}$ are then constants.
By taking only the integrable low order terms into account,
we can write down an asymptotic
formula for the shear stress:
\begin{equation}
\sigma=\sum_{n=0}^{\left\lfloor \frac{2}{c}\right\rfloor _{-}}\sigma_{n}^{*}\left(a_{1}\left|\dot{\gamma}\right|^{c}\right)^{n}.
\label{eq:sigma_approx}
\end{equation}
Neglected higher order terms  ${\cal O}\left(\left|a_1\right|^{\frac{2}{c}}\left|\dot{\gamma}\right|^2\right) $ are subdominant.
Now we use the two-parameter scaling law for $a_{1}$ with
the choice $\Omega=\left|\dot{\gamma}t_{0}\right|^{\frac{1}{1+a}}$
and define
\begin{equation}
\Lambda\left(x\right)=\frac{\hat{a}_{1}\left(x,1\right)}{\hat{a}_{1}\left(0,1\right)}
\label{eq:def_lambda}
\end{equation}
for some arbitrary argument $x$. With the definitions $\sigma_{n}^{0}=\left(\hat{a}_{1}\left(0,1\right)\right)^{n}\sigma_{n}^{*}$,
\begin{equation}
\tilde{m}=\frac{2a}{1+a},
\end{equation}
\begin{equation}
m'=\frac{c}{2},
\end{equation}
\begin{equation}
m=\tilde{m}m',
\end{equation}
we obtain the $\Lambda$-formula:
\begin{equation}
\sigma\left(\varepsilon,\dot{\gamma}\right)=\sum_{n=0}^{\left\lfloor \frac{1}{m'}\right\rfloor _{-}}\sigma_{n}^{0}\left(\Lambda\left(\frac{\varepsilon}{\varepsilon_{\dot{\gamma}}}\right)\left|\dot{\gamma}t_{0}\right|^{m}\right)^{n},
\label{eq:lambda_formula}
\end{equation}
\begin{equation}
\varepsilon_{\dot{\gamma}}=\left|\dot{\gamma}t_{0}\right|^{\tilde{m}}.
\label{eq:eps_gammadot2}
\end{equation}
This is the central result of our work. Eq. (\ref{eq:lambda_formula}) generalizes the results presented in \cite{Fuc:03} and \cite{Hen:05} and summarizes many aspects of the flow curves.

Eq. (\ref{eq:lambda_formula}) was obtained by integrating the correlators obeying the generalized 'time-shear-superposition principle'. Their amplitudes depend sensitively on shear rate and separation from the glass transition, as could be determined from the glass stability analysis. One function $\Lambda\left(\varepsilon/\varepsilon_{\dot\gamma}\right)= \hat{a}_1\left(\varepsilon/\varepsilon_{\dot\gamma},1\right)/\hat{a}_{1}\left(0,1\right)$ remained capturing this sensitive dependence, and in turn now determines the flow curves.
In the following we discuss the properties of the $\Lambda$-formula
in detail. General requirements for the range of validity are $\left|\varepsilon\right|\ll1$
and $\left|\dot{\gamma}t_{0}\right|\ll1$. The $\sigma_{n}^{0}$ are constants, which for our
model numerically are obtained from fits.

\subsection{The $\Lambda$-function}

\begin{figure}
\resizebox{1\columnwidth}{!}{\includegraphics{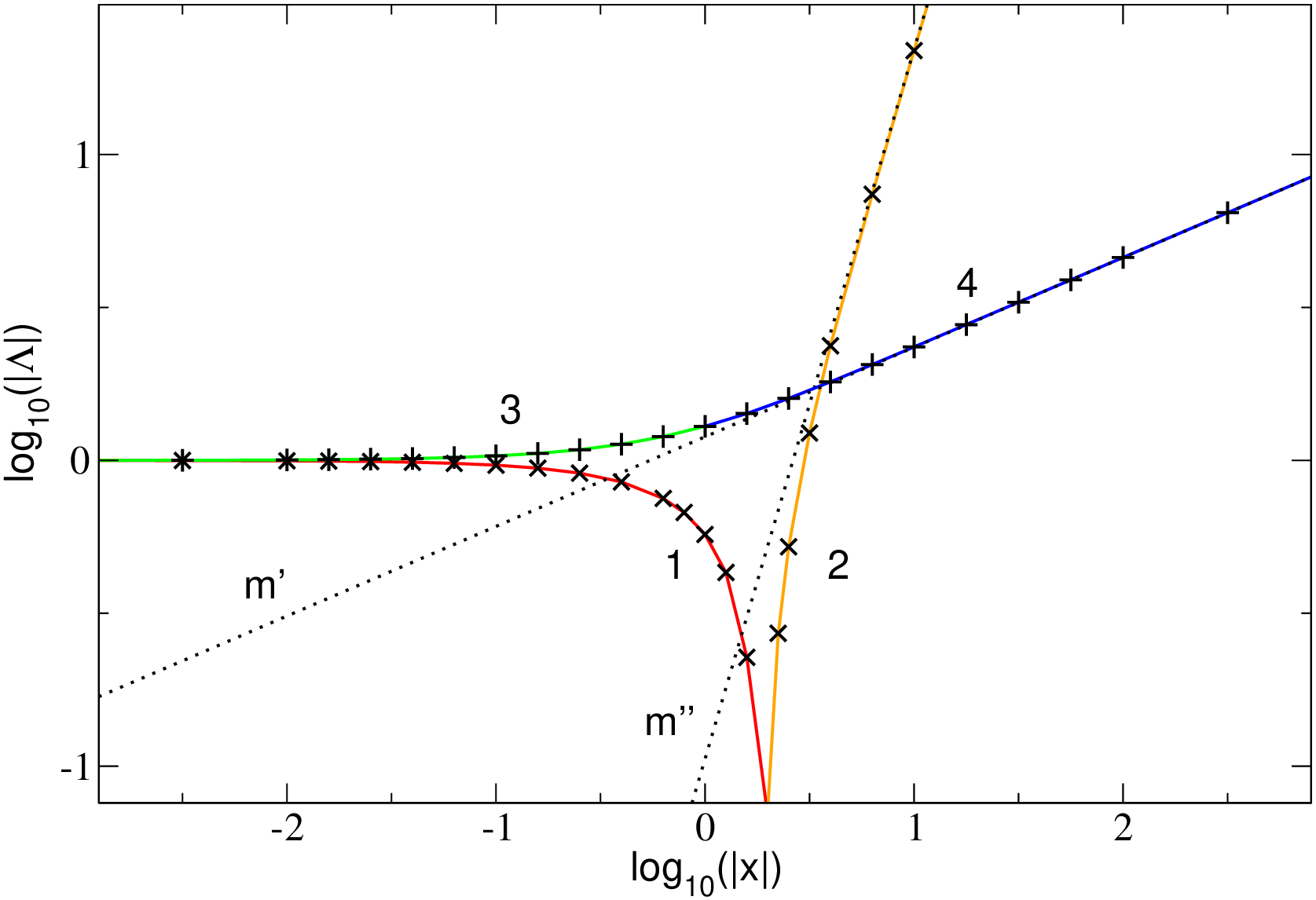}}
\caption{\label{fig:3} Numerically obtained data for $\Lambda\left(x\right)$
(crosses for $x<0$, plus signs for $x>0$) and the fitted functions $\Lambda_{<}^{-}\left(x\right)$ for small negative $x$
(red, 1), $\Lambda_{>}^{-}\left(x\right)$ for large negative $x$ (orange, 2), $\Lambda_{<}^{+}\left(x\right)$ for small positive $x$
(green, 3) and $\Lambda_{>}^{+}\left(x\right)$ for large positive $x$ (blue, 4), see Appendix \ref{sec:lambda} for details. With these functions
$\Lambda\left(x\right)$ can be approximated for arbitrary arguments
with sufficiently high accuracy. The dotted lines show
the leading asymptotes with the exponents $m''$ and $m'$.}
\end{figure}

The central aspect of our work was to analyze the shape of $\Lambda\left(x\right)$ which we can determine
numerically from $\mathcal{G}\left(t\right)$. This scalar function
of one real parameter determines the shape of the flow curves within the range of validity of the $\Lambda$-formula.
For instance, the asymptote for $x\rightarrow-\infty$ describes the flow curves in the liquid region while the
asymptote for $x\rightarrow\infty$ describes the scaling of the yield stress for $\varepsilon>0$.
We can determine three limiting behaviors.

By using Eq. (\ref{eq:a1_liquid}), Eq. (\ref{eq:a1_transition}) and
Eq. (\ref{eq:a_1_glass}) we identify
\begin{equation}
\Lambda\left(x\rightarrow-\infty\right)=
-\Lambda_{\infty}^{-}\left|x\right|^{m''},
\label{eq:lambda_liquid}
\end{equation}
\begin{equation}
m''=\frac{\gamma bc}{1-b},
\end{equation}
\begin{equation}
\Lambda\left(0\right)=1,
\label{eq:lambda_trans}
\end{equation}
\begin{equation}
\Lambda\left(x\rightarrow\infty\right)=\Lambda_{\infty}^{+}x^{m'}.
\label{eq:lambda_glass}
\end{equation}
We determine $\Lambda_{\infty}^{-}$ and $\Lambda_{\infty}^{+}$ numerically.
Fig. \ref{fig:3} shows an overview of the properties of the $\Lambda$-function. A more detailed discussion can be found
in Appendix \ref{sec:lambda}.

\subsection{Discussion of the flow curves}

\begin{figure}
\resizebox{1\columnwidth}{!}{\includegraphics{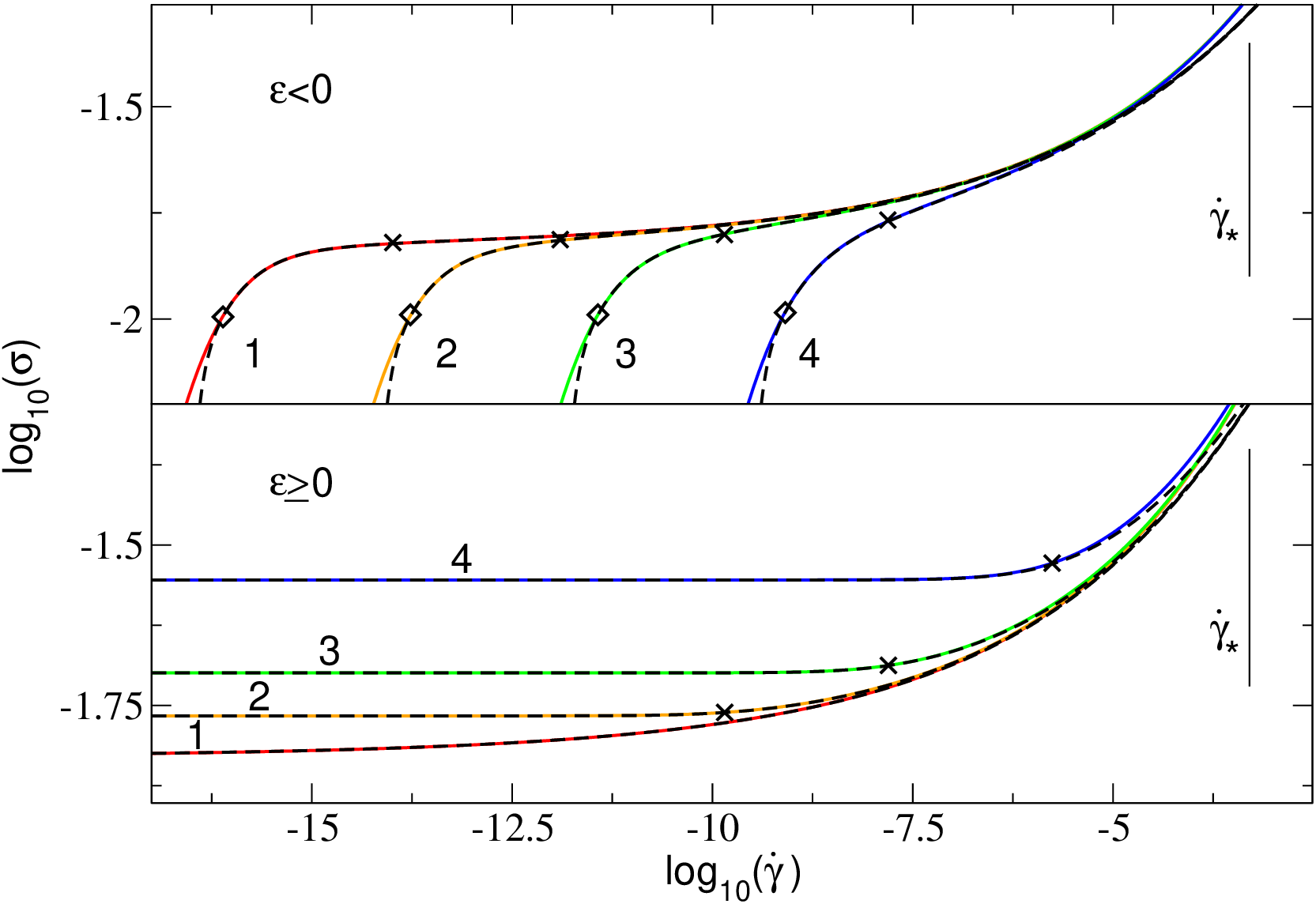}}
\caption{\label{fig:5} Overview of the numerically obtained flow curves (solid lines) and
the $\Lambda$-formula evaluated numerically (dashed lines). The liquid
curves in the upper panel are shown for $\varepsilon=-10^{-7}$ (red, 1), $\varepsilon=-10^{-6}$ (orange, 2), $\varepsilon=-10^{-5}$
(green, 3) and $\varepsilon=-10^{-4}$ (blue, 4). The lower panel shows
the glassy curves for $\varepsilon=0$ (red, 1), $\varepsilon=10^{-5}$ (orange, 2), $\varepsilon=10^{-4}$ (green, 3)
and $\varepsilon=10^{-3}$ (blue, 4). Crosses mark the points with $\left|\varepsilon\right|=\varepsilon_{\dot{\gamma}}$.
The natural upper boundary for the shear rate,
where the range of validity of the $\Lambda$-formula is left, is also indicated, see Eq. (\ref{eq:crit_flow_2}).
For $\varepsilon<0$, the natural lower limits for the shear rates described by Eq. (\ref{eq:lower_limit}), below which the $\Lambda$-formula does not describe the flow curves, are marked by diamonds.}
\end{figure}

Fig. \ref{fig:5} shows an overview of the numerically obtained flow
curves and the corresponding numerical results for the
$\Lambda$-formula. While the glass flow curves exhibit an upward
curvature only, the fluid curves show a characteristic S-shape,
where the initial downward curvature changes to an upward one for
increasing  shear rate. For positive separation parameters the range
of validity of the $\Lambda$-formula is given by
$\left|\varepsilon\right|\ll1$ and
$\left|\dot{\gamma}t_{0}\right|\ll1$. These two requirements ensure
that $\mathcal{G}\left(t\right)$ describes the dynamics of
$\Phi\left(t\right)$ with a sufficiently high accuracy. For
sufficiently small negative separation parameters the
$\Lambda$-formula is valid in finite shear rate windows only, as it
does not reproduce the linear asymptotes for low shear rates. We
find a precise criterion for the range of validity later when we
discuss the limiting cases.

In the framework of the $\Lambda$-formula, it is easy to understand
the mathematical structure of the singular behavior of the yield
stress at $\varepsilon=0$: We choose some fixed $\varepsilon=\pm\left|\varepsilon\right|$
and consider the limit $\dot{\gamma}\rightarrow0$. Then for $x=\varepsilon/\varepsilon_{\dot{\gamma}}$
we obtain $x\rightarrow\pm\infty$. The singular behavior is then
due to the different asymptotes of $\Lambda\left(x\right)$ for $x\rightarrow\pm\infty$
while $\Lambda\left(x\right)$ itself is a smooth function for all
$x$.

While Fig. \ref{fig:5} presents an overview of the flow curves, detailed discussions of the various limits are presented next.

\subsubsection{Yield stress}

\begin{figure}
\resizebox{1\columnwidth}{!}{\includegraphics{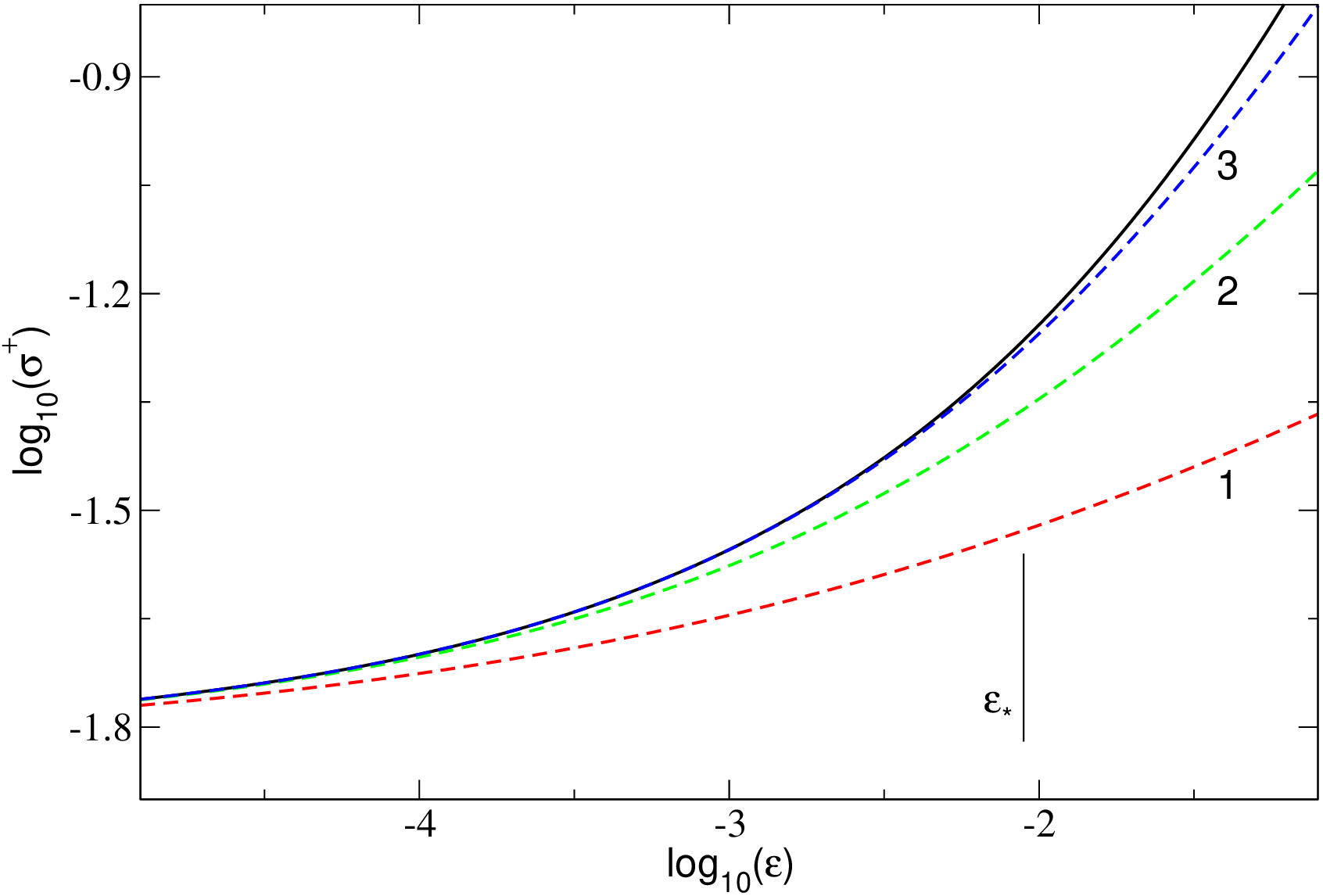}}
\caption{\label{fig:6a} Numerically obtained curve for the
yield stress (solid line). The dashed lines show the corresponding power series given by Eq.
(\ref{eq:yield_stress}) evaluated to first order (red, 1), second order
(green, 2) and third order (blue, 3). The natural upper boundary for the range of validity of Eq. (\ref{eq:yield_stress}), respectively
Eq. (\ref{eq:yield_stress_2}), is also indicated.}
\end{figure}

By using Eq. (\ref{eq:lambda_glass}) we obtain an asymptotic formula
for the yield stress which reproduces a result in \cite{Fuc:03}:
\begin{equation}
\sigma^{+}\left(\varepsilon\right)=\sum_{n=0}^{\left\lfloor \frac{1}{m'}\right\rfloor _{-}}\sigma_{n}^{+}\varepsilon^{m'n}.
\label{eq:yield_stress}
\end{equation}
The constants are defined by $\sigma_{n}^{+}=\sigma_{n}^{0}\left(\Lambda_{\infty}^{+}\right)^{n}$.
We determine $\sigma_{0}^{+}=\sigma^{+}\left(\varepsilon=0\right)$
by extrapolating the flow curve for $\varepsilon=0$. Then we use
a non-linear curve fitting to determine $\sigma_{n\geq1}^{+}$.  The quantity $\sigma_{0}^{+}$ is the dynamic yield stress at the transition, also denoted by $ \sigma_{c}^{+}$; numerically we verify that it agrees with the integral
$ \sigma_{c}^{+}=\tilde{\dot{\gamma}} \int_0^{\infty}  \tilde \Phi_0^2(\tilde t) d\tilde t$, as implied by the time-shear superposition principle Eq.   (\ref{eq:alpha_series}). Equation (\ref{eq:yield_stress})
describes the yield stress for sufficiently small positive separation
parameters, see Fig. \ref{fig:6a}, and
%Eq. (\ref{eq:yield_stress})
can be rewritten as
\begin{equation}
\sigma^{+}\left(\varepsilon\right)=\sigma_{0}^{+}\sum_{n=0}^{\left\lfloor \frac{1}{m'}\right\rfloor _{-}}c_{n}^{+}\left(\varepsilon/\varepsilon_{*}\right)^{m'n}
\label{eq:yield_stress_2}
\end{equation}
with $c_{0}^{+}=1$. With the choice $c_{1}^{+}=1$ we can identify
$\varepsilon_{*}=\left(\sigma_{0}^{+}/\sigma_{1}^{+}\right)^{\frac{1}{m'}}$ as a natural scale for the separation parameter in our
asymptotic expansion. To keep inside the range of validity of the $\Lambda$-formula we have to require
$\left|\varepsilon\right|\ll\varepsilon_{*}$.
For the constants we obtain $c_{n\geq2}^{+}=\left(\sigma_{n}^{+}/\sigma_{0}^{+}\right)\varepsilon_{*}^{m'n}$.

\subsubsection{Critical curve}

\begin{figure}
\resizebox{1\columnwidth}{!}{\includegraphics{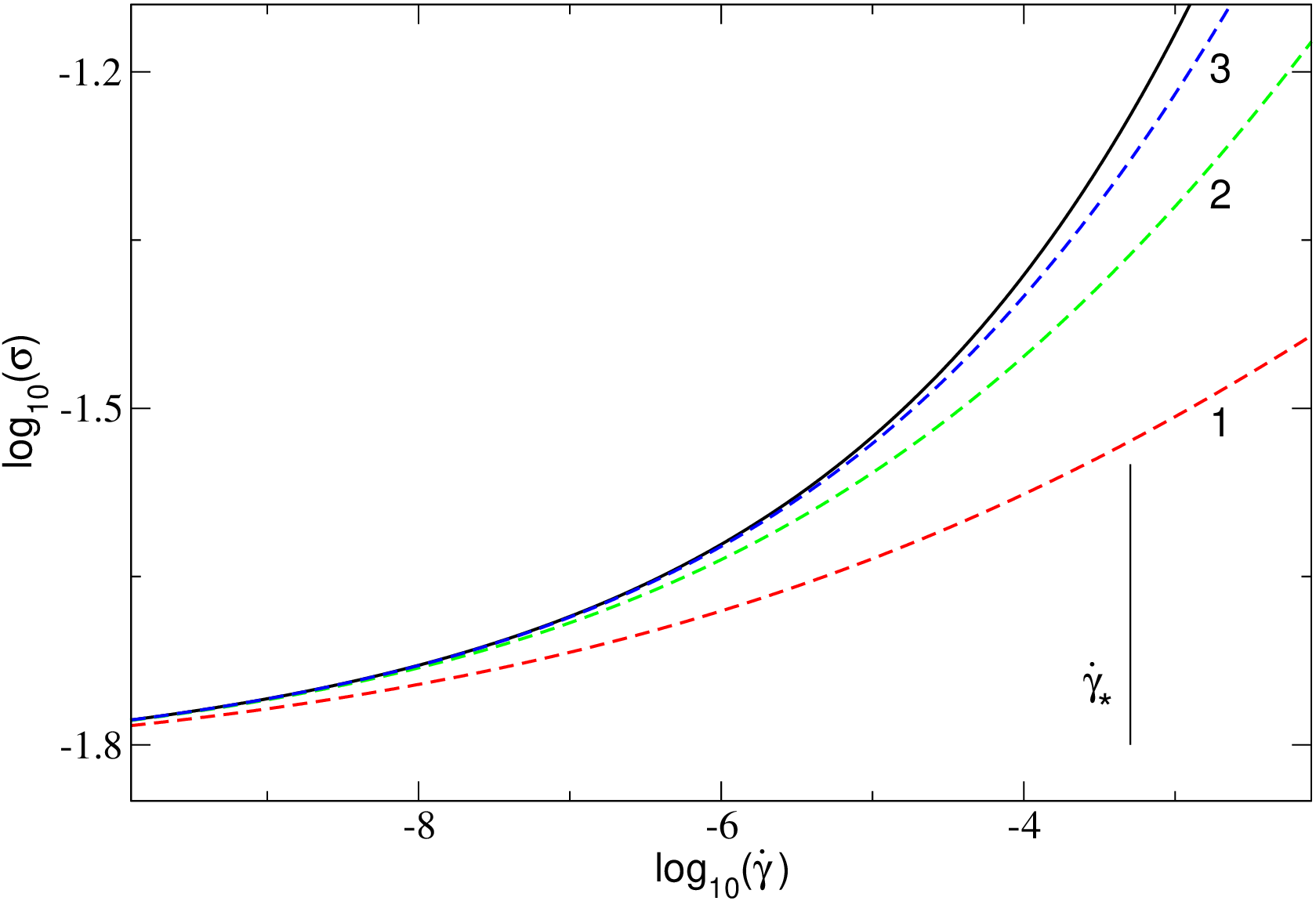}}
\caption{\label{fig:6b} Numerically obtained flow curve for $\varepsilon=0$ (solid line).
The dashed lines show the corresponding power series given by Eq. (\ref{eq:crit_flow})
evaluated to first order (red, 1), second order
(green, 2) and third order (blue, 3). The natural upper boundary for the range of validity of Eq. (\ref{eq:crit_flow}), respectively
Eq. (\ref{eq:crit_flow_2}), is also indicated.}
\end{figure}

By using Eq. (\ref{eq:lambda_trans}) we obtain an asymptotic formula
for the critical flow curve reproducing the results shown in \cite{Fuc:03} and \cite{Hen:05}, the generalized Herschel-Bulkley law:
\begin{equation}
\sigma\left(\varepsilon=0,\dot{\gamma}\right)=\sum_{n=0}^{\left\lfloor \frac{1}{m'}\right\rfloor _{-}}\sigma_{n}^{0}\left|\dot{\gamma}t_{0}\right|^{mn}.
\label{eq:crit_flow}
\end{equation}
We determine the constants directly from the fitted values
for $\sigma_{n}^{+}$ by using $\sigma_{n}^{+}=\sigma_{n}^{0}\left(\Lambda_{\infty}^{+}\right)^{n}$.
We observe that for $\varepsilon=0$
Eq. (\ref{eq:crit_flow}) describes the flow curve correctly for
sufficiently small shear rates, see Fig. \ref{fig:6b}.
Eq. (\ref{eq:crit_flow}) can be rewritten as
\begin{equation}
\sigma\left(\varepsilon=0,\dot{\gamma}\right)=\sigma_{c}^{+}\sum_{n=0}^{\left\lfloor \frac{1}{m'}\right\rfloor _{-}}c_{n}\left|\dot{\gamma}/\dot{\gamma}_{*}\right|^{mn}.
\label{eq:crit_flow_2}
\end{equation}
With the choice $c_{0}=c_{1}=1$ we can identify $\sigma_{c}^{+}=\sigma_{0}^{0}$ as the critical dynamic yield stress
and $\dot{\gamma}_{*}=\left(t_{0}\right)^{-1}\left(\sigma_{0}^{0}/\sigma_{1}^{0}\right)^{\frac{1}{m}}$ defines a natural scale for the shear rates in our asymptotic expansion \cite{Cra:07}. To keep
inside the range of validity of the $\Lambda$-formula we have to require $\left|\dot{\gamma}\right|\ll\dot{\gamma}_{*}$.
For the constants we obtain $c_{n\geq2}=\left(\sigma_{n}^{0}/\sigma_{0}^{0}\right)\left(\dot{\gamma}_{*}t_{0}\right)^{mn}$.

\subsubsection{Liquid}

\begin{figure}
\resizebox{1\columnwidth}{!}{\includegraphics{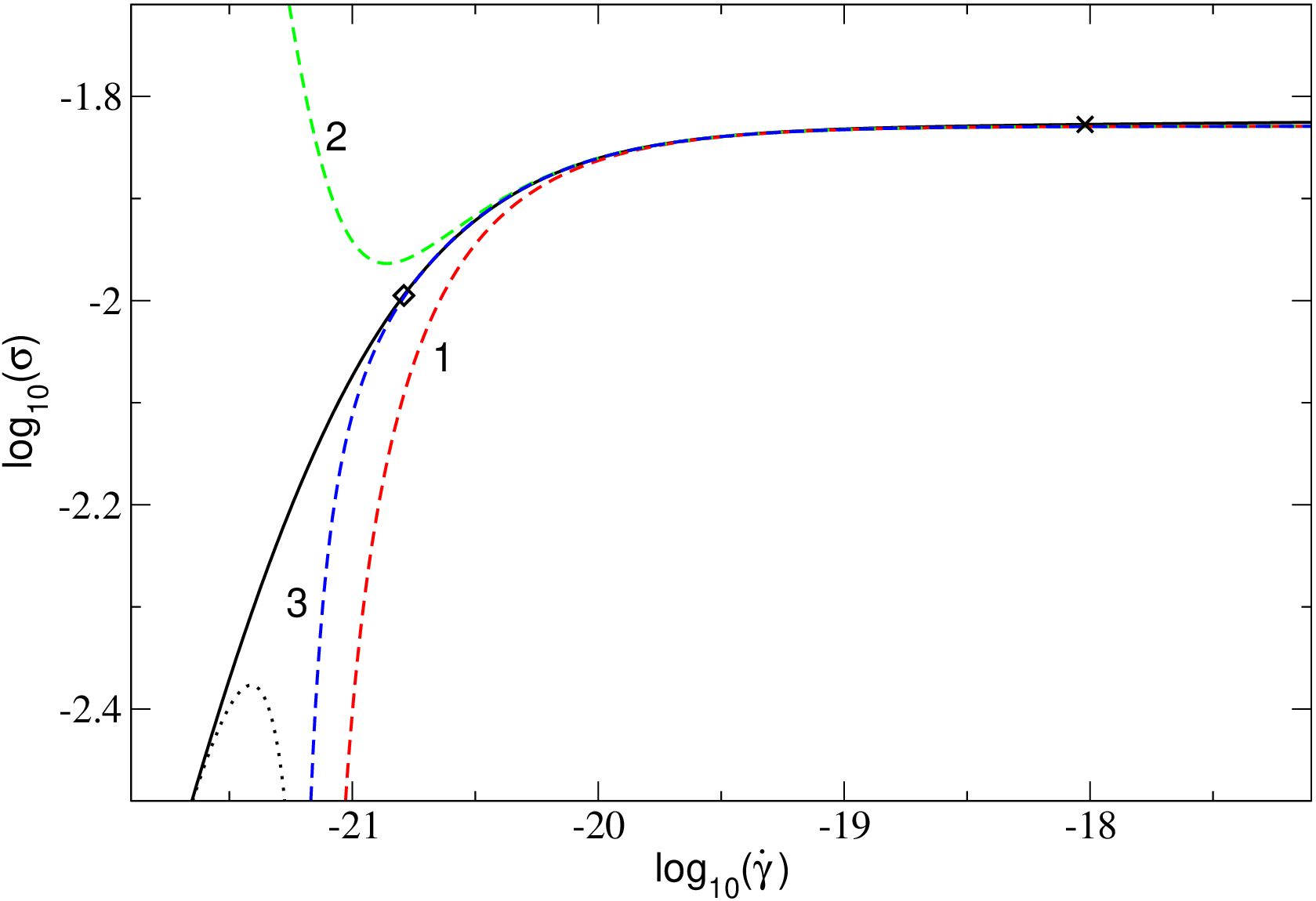}}
\caption{\label{fig:6c} Numerically obtained flow curve for $\varepsilon=-10^{-9}$ (solid line).
The dashed lines show the corresponding power series given by Eq. (\ref{eq:formula_luquid}) evaluated to first order (red, 1),
second order (green, 2) and third order (blue, 3). The shear rate
with $\varepsilon=-\varepsilon_{\dot{\gamma}}$ is marked by a cross.
The natural lower limit $\dot{\gamma}_{min}$ for the shear rate defined by Eq. (\ref{eq:lower_limit}) is marked by a diamond.
A fitted regular Taylor expansion to the order $\dot{\gamma}^{3}$ for the region below $\dot{\gamma}_{min}$ is also included (dotted line).
We remark that we need extremely small $\varepsilon$
and $\dot{\gamma}$ to verify Eq. (\ref{eq:formula_luquid}) numerically.}
\end{figure}

Eq. (\ref{eq:lambda_liquid}) leads us to a formula for negative separation
parameters:
\begin{equation}
\sigma\left(\varepsilon\ll-\varepsilon_{\dot{\gamma}},\dot{\gamma}\right)=\sum_{n=0}^{\left\lfloor \frac{1}{m'}\right\rfloor _{-}}\left(-1\right)^{n}\sigma_{n}^{-}\left|\varepsilon\right|^{m''n}\left|\dot{\gamma}t_{0}\right|^{-\bar{m}n},
\label{eq:formula_luquid}
\end{equation}
\begin{equation}
\bar{m}=\frac{m''}{\gamma}.
\end{equation}
We determine the constants
directly from the values for $\sigma_{0}^{0}$ by evaluating $\sigma_{n}^{-}=\sigma_{n}^{0}\left(\Lambda_{\infty}^{-}\right)^{n}$.
We observe that for sufficiently small
separation parameters Eq. (\ref{eq:formula_luquid}) describes the
flow curves correctly in finite shear rate windows where the flow
curves are located close below the critical yield stress value, see Fig. \ref{fig:6c}.

A necessary criterion for the validity of Eq. (\ref{eq:formula_luquid})
is that the matching time $\tau_{b}\propto\left|\varepsilon\right|^{\gamma\frac{b}{1-b}}\left|\dot{\gamma}t_{0}\right|^{-\frac{1}{1-b}}$
lies inside the range of validity of the nonlinear stability equation, see Eq. (\ref{eq:beta_scaling}).
This criterion is equivalent to the statement that $\mathcal{G}\left(t\right)$
converges to the long time asymptote $-t/\tau_{\dot{\gamma}}$ under
the condition $\left|\mathcal{G}\left(t\right)\right|\ll1$.
This consideration leads to lower limits for the shear rates, below which the $\Lambda$-formula does not describe the flow curves. They are marked by diamonds in Fig. \ref{fig:5} and Fig. \ref{fig:6c}. The requirement $\tau_{b}\ll\tau_{\dot{\gamma}}$ leads to $\left|\dot{\gamma}\right|\gg\dot{\gamma}_{min}$ with
\begin{equation}
\dot{\gamma}_{min}=\left(\hat{\tau}_{0}t_{0}\right)^{-1}\left(\tilde{\dot{\gamma}}\right)^{-\frac{1-b}{b}}\left|\varepsilon\right|^{\gamma}.
\label{eq:lower_limit}
\end{equation}
Below this limit, a regular Taylor expansion holds which is also shown in Fig. \ref{fig:6c}.

\subsubsection{Yielding glass}

\begin{figure}
\resizebox{1\columnwidth}{!}{\includegraphics{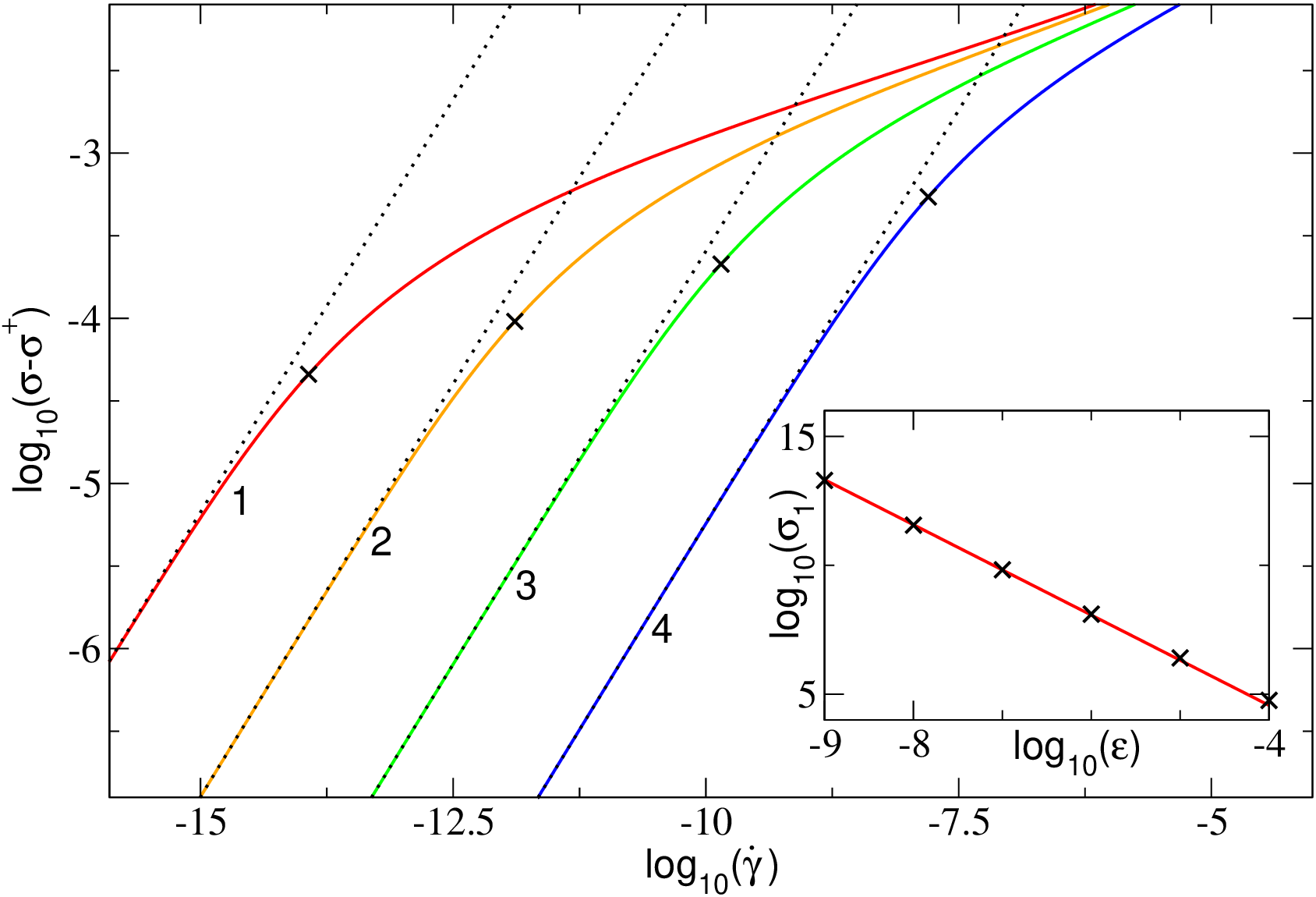}}
\caption{\label{fig:7} Numerically obtained flow curves
from which the yield stress is subtracted (solid lines) for $\varepsilon=10^{-7}$
(red, 1), $\varepsilon=10^{-6}$ (orange, 2), $\varepsilon=10^{-5}$
(green, 3) and $\varepsilon=10^{-4}$ (blue, 4).
Crosses mark the points with $\varepsilon=\varepsilon_{\dot{\gamma}}$.
The dotted lines show the fitted linear functions $\sigma_{1}\left|\dot{\gamma}\right|$. The inset
demonstrates that the fitted coefficients (crosses) obey $\sigma_{1}\propto\varepsilon^{-\zeta}$
(solid line) for sufficiently small $\varepsilon$. }
\end{figure}

Another interesting aspect is that for $\varepsilon>0$ we observe numerically a linear regime for small shear rates:
\begin{equation}
\sigma\left(\varepsilon>0,\dot{\gamma}\rightarrow0\right)=\sigma^{+}+\sigma_{1}\left|\dot{\gamma}\right|.
\label{eq:sigma_linear}
\end{equation}
This result enables us to identify the next-to-leading order asymptote of the $\Lambda$-function for large positive arguments,
see Appendix \ref{sec:lambda}. With this we can predict a scaling law for the coefficient $\sigma_1$ for small $\varepsilon$:
\begin{equation}
\sigma_{1} \propto \varepsilon^{-\zeta},
\label{eq:linear_scaling}
\end{equation}
\begin{equation}
\zeta=\frac{1-m}{\tilde{m}}.
\end{equation}
These formulae are consistent with our numerical results shown in Fig. \ref{fig:7}.
Note that Eq. (\ref{eq:sigma_linear}) defines a function which is not analytic in $\dot{\gamma}$.

\subsection{The pseudo power law}

\begin{figure}
\resizebox{1\columnwidth}{!}{\includegraphics{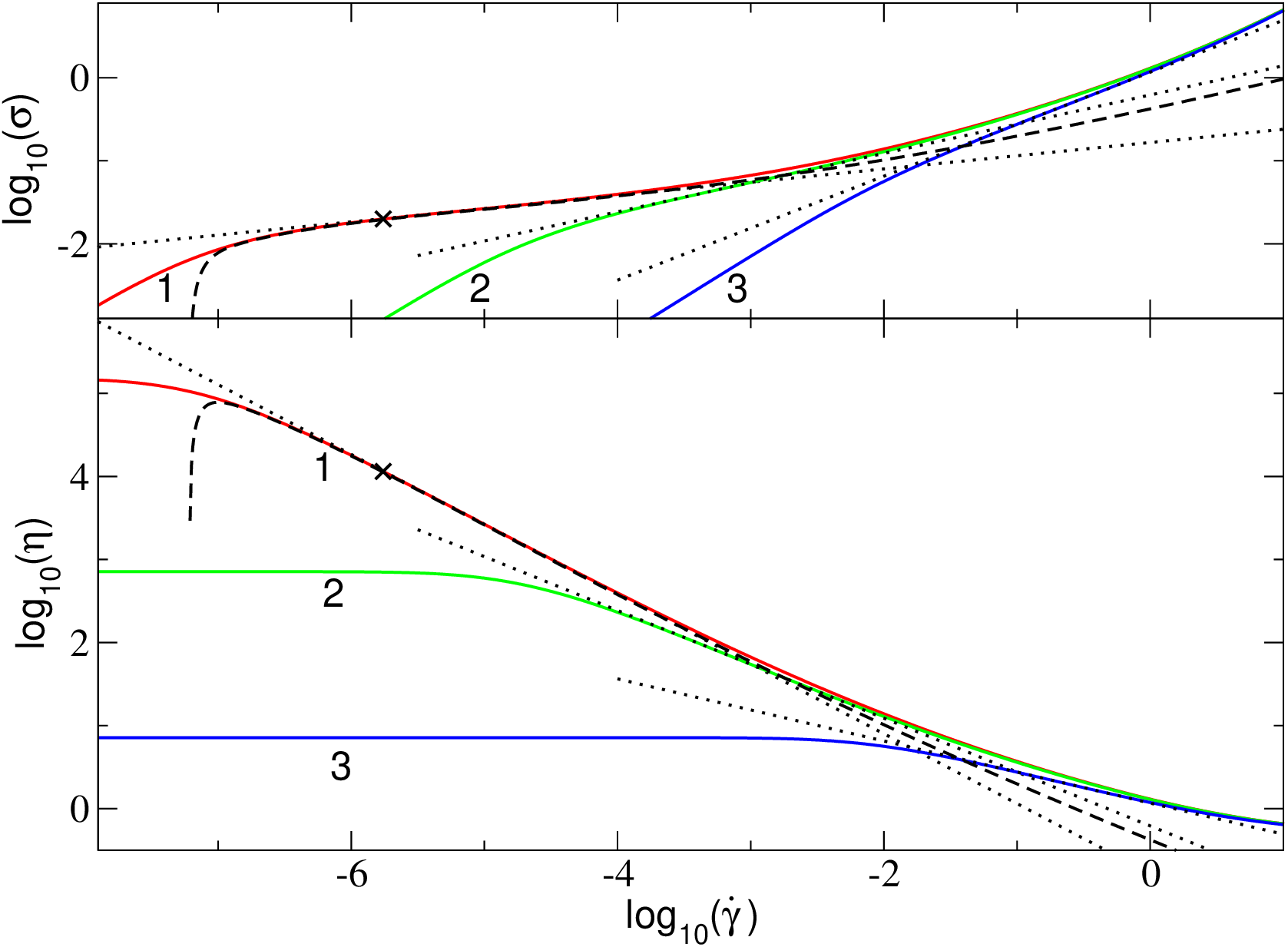}}
\caption{\label{fig:8} The upper panel shows numerically obtained flow curves (solid lines) for $\varepsilon=-10^{-3}$ (red, 1), $\varepsilon=-10^{-2}$ (green, 2) and $\varepsilon=-10^{-1}$ (blue, 3). The dotted lines show the corresponding inflection tangents, with exponents $p=0.16$, $0.35$, and $0.63$ from left to right.
The dashed line shows the numerically evaluated $\Lambda$-formula
for $\varepsilon=-10^{-3}$. The shear rate with $\varepsilon=-\varepsilon_{\dot{\gamma}}$ is marked
by a cross. The lower panel shows the corresponding results for the viscosity.}
\end{figure}

For $\varepsilon<0$ we observe that $\log_{10}\left(\sigma\right)$
as a function of $\log_{10}\left(\dot{\gamma}\right)$ shows an inflection
point defined by
\begin{equation}
\frac{d^{2}\left(\log_{10}\left(\sigma\right)\right)}{d\left(\log_{10}\left(\dot{\gamma}\right)\right)^{2}}=0.
\label{eq:inflection_point}
\end{equation}
Obviously, this statement is also true for any (toy-) model
which shows shear-thinning. But then in some finite shear rate windows
the flow curves can be approximated by the corresponding inflection
tangents. The slopes $p$ of the inflection tangents can be interpreted
as exponents occurring in some pseudo power laws:
\begin{equation}
\sigma \propto \dot{\gamma}^{p}.\label{eq:pseudo_power}
\end{equation}
Fig. \ref{fig:8} shows some examples. We observe that the
numerically evaluated $\Lambda$-formula also describes the
neighborhood of the inflection point correctly for sufficiently
small $\varepsilon$. As shown in the plot, in this region
$-\varepsilon_{\dot{\gamma}}<\varepsilon<0$ is satisfied, hence the
$\Lambda$-function can be described by Eq. (\ref{eq:lambda_5}), see
Appendix \ref{sec:lambda} for details. By substituting Eq.
(\ref{eq:lambda_5}) in Eq. (\ref{eq:lambda_formula}) we see that the
resulting formula does not represent a real power law of the type of
Eq. (\ref{eq:pseudo_power}).

We conclude that, in the framework of our asymptotic expansion,
there is no real exponent $p$. Eq. (\ref{eq:pseudo_power}) is a
trivial artifact of the double logarithmic plot of the flow curves
which then necessarily show inflection points for any model
describing shear thinning. Rather, the flow curves on the fluid side
exhibit a characteristic S-shape. While this shape is rather
apparent when plotting stress as function of shear rate, plotting
the same data as viscosity as function of shear rate hides it. The
vertical axis gets appreciably stretched, and ruling out the pseudo
power law becomes more difficult using experimental data.

\section{Flow curves: intermediate regime}

The correlator $\tilde \Phi(\tilde t)$ from Eq.  (\ref{eq:rescaled_mori})
determines the structural contribution to the stationary stress,
where microscopic, initial-decay-rate-$\Gamma$-dependent
contributions are neglected. It contains the complete structural
dynamics on time scales longer than $t_0$, and starts out with the critical power law  according to Eqs.~(\ref{eq:factorization}) and (\ref{eq:critical_decay}). Expansion (\ref{eq:alpha_series}) captures its behavior for rescaled
times $\tilde t \gg \sqrt{\varepsilon_{\dot \gamma}}$ only, because
the power series (\ref{eq:long_time_expansion}), on which expansion (\ref{eq:alpha_series}) is based, at this
scale crosses over to expansions on shorter time scales; the
matchings are discussed in Sect. 3.4. The divergence of the
expansion coefficients $\tilde \Phi_n(\tilde t)$ for $\tilde t\to0$
thus is only apparent, and becomes cut off at $\tilde t = {\cal O}(
\sqrt{\varepsilon_{\dot \gamma}})$.

\begin{figure}
\resizebox{1\columnwidth}{!}{\includegraphics{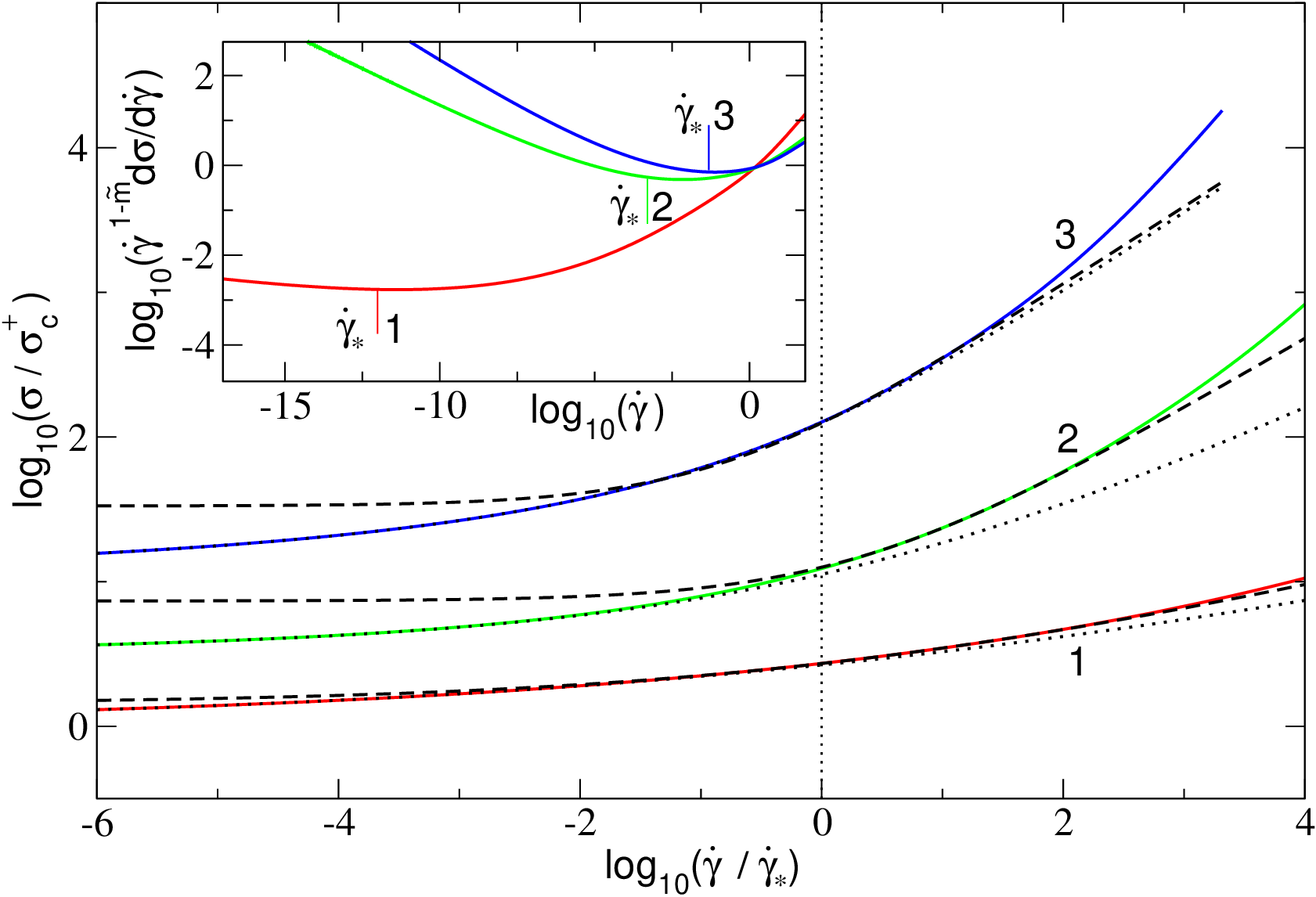}}
\caption{\label{fighb} Critical flow curves for three different transitions and exponent parameters $\lambda$ rescaled using the critical yield stress $\sigma^+_c$ and the scale $\dot\gamma_*$ to agree with the Herschel-Bulkley law from Eq.~(\ref{neu4}) with exponents $\tilde m=0.200 $ ($\lambda=0.976$, red, label 1), $\tilde m=0.489 $ ($\lambda=0.707$, green, label 2), and $\tilde m=0.542 $ ($\lambda=0.577$, blue, label 3); curves 2 and 3 are shifted upwards by $\frac 12$ and 1, respectively. Solid lines are the numerical flow curves, dotted lines the generalized Herschel-Bulkley law from Eq.~(\ref{eq:crit_flow_2}) which holds for $\dot\gamma<\dot\gamma_*$, and dashed lines give Eq.~(\ref{neu4}) which holds for $\dot\gamma\sim\dot\gamma_*$. The inset shows the multiplied derivatives $\dot \gamma^{1-\tilde m} d\sigma/d\dot\gamma$, where the corresponding scales $\dot\gamma_*$ are marked.
}
\end{figure}

 Let us define a correlator
$\doubletilde \Phi(\tilde t)$, where the (dominant) divergent
initial variations are subtracted:
\begin{equation}\label{neu1}
\doubletilde \Phi(\tilde t)=\tilde \Phi(\tilde t)   -
(1-f_c)^2\; \sqrt{\varepsilon_{\dot \gamma}}\; {\cal G}(\tilde t
/\sqrt{\varepsilon_{\dot \gamma}}) \; \Theta( 1-\tilde t)  \; ,
\end{equation}
where $\cal G$ is the series from Eq.~(\ref{eq:long_time_expansion}) valid for
$\sqrt{\varepsilon_{\dot \gamma}}\ll \tilde t \ll 1$, only. The new
correlator possesses an expansion like Eq.~(\ref{eq:alpha_series}) with coefficient
functions $\doubletilde \Phi_n(\tilde t)$ that behave more regularly at short rescaled times. It
thus can be integrated over rescaled time $\tilde t$. The
Heaviside-step function in Eq.~(\ref{neu1}) serves as remembrance
that the series (\ref{eq:long_time_expansion}) only describes the
initial variation of the yield scaling functions $\tilde \Phi$. The
sum,
\begin{equation}\label{neu2} \tilde \Phi(\tilde t)= \doubletilde
\Phi(\tilde t)   + (1-f_c)^2\; \sqrt{\varepsilon_{\dot \gamma}}\;
{\cal G}(\tilde t /\sqrt{\varepsilon_{\dot \gamma}}) \; \Theta(
1-\tilde t)  \; ,
\end{equation}
where now the complete $\beta$-correlator $\cal G$ from Eq. (\ref{eq:beta_scaling}) is
taken, thus describes the solution of Eq. (\ref{eq:rescaled_mori}) down to its limit of
validity, $\left|\dot\gamma t_0\right| \ll {\tilde t}$.

In the limit of small shear rates, $\left|\dot \gamma t_0\right| \ll 1$, the
stress from Eq.~(\ref{eq:shear_stress}) using this rewriting of $\tilde \Phi$ becomes:
\begin{eqnarray}\label{neu3} \nonumber
\sigma &=&  \tilde{\dot\gamma}\; \int_0^\infty  \left( \doubletilde
\Phi(\tilde t)   + (1-f_c)^2\; \sqrt{\varepsilon_{\dot \gamma}}\;
{\cal G}(\tilde t /\sqrt{\varepsilon_{\dot \gamma}}) \; \Theta(
1-\tilde t) \right)^2 d\tilde t\;\\
 &=& \tilde{\dot\gamma}\; \int_0^\infty\!\!\!  \doubletilde
\Phi^2(\tilde t) d\tilde t\;  + h_\sigma\; \varepsilon_{\dot \gamma}\; \int_0^{1/\sqrt{\varepsilon_{\dot \gamma}}}\!\!\!  {\cal G}(\hat t )d\hat t\;  + {\cal O}(\varepsilon_{\dot \gamma}^{3/2})
\end{eqnarray} where $\hat
t= \tilde t /\sqrt{\varepsilon_{\dot \gamma}}$; the constant abbreviates
 $h_\sigma=2\tilde{\dot\gamma}\,f_c\, (1-f_c)^2$.

The  result just obtained agrees with Eq.~(\ref{eq:crit_flow_2}) in the asymptotic
limit $\dot \gamma\to0$. This holds because the scaling function
${\cal G}(\hat t)$ diverges for long rescaled times $\hat t$, and
thus the behavior of the second integral in Eq. (\ref{neu3}) at the
upper limit, which shifts to infinity in this case, gives the
leading contributions. Yet, the scaling function $\cal G$ remains
close to zero for rescaled times $\hat t$, which  correspond to the
region where the full correlator stays close to the plateau $f_c$;
see Fig.~\ref{fig:fig_1}. For not too small shear rates, which we estimate based
on  Eq.~(\ref{eq:crit_flow_2}) by $\left|\dot \gamma \right| \gtrapprox \dot \gamma_*$, the second integral
in Eq.~(\ref{neu3}) thus does not depend on $\dot \gamma$, and we
derive a Herschel-Bulkley law for intermediate shear rates:
\begin{equation}
\label{neu4}
\sigma(\dot \gamma_*\ll \left|\dot \gamma\right| \ll \Gamma,\varepsilon=0 )  =
\tilde \sigma_0 + \tilde \sigma_1
\left|\dot \gamma t_0 \right|^{\tilde m} + {\cal O}(\left|\dot \gamma t_0 \right|^{3\tilde m/2})
\end{equation}
with constant amplitudes
$$ \tilde \sigma_0 =\tilde{\dot{\gamma}}\int_0^\infty
\doubletilde \Phi_0^2(\tilde t)\;d\tilde t\; ,\;
\mbox{and }\; \tilde \sigma_1 = h_\sigma\,
\int_0^{1/\sqrt{\varepsilon_{\dot \gamma_*}}} {\cal
G}(\hat t)\; d\hat t\; .$$
The constant $\tilde \sigma_0$ is not the actual yield stress,
$\sigma^+$, which is obtained in the limit of vanishing shear rate,
$\sigma^+=\sigma(\dot\gamma\to0)$, but is larger, because $\doubletilde
 \Phi_0(\tilde t)$ remains longer close to the plateau value
$f_c$ than $\tilde \Phi_0(\tilde t)$, which enters in the definition
of $\sigma^+$; this follows because $\doubletilde \Phi_0(\tilde t)$
lacks the initial term linear in $\tilde t$.

Fig. \ref{fighb} verifies the Herschel-Bulkley law for three
different exponents $\tilde m=\frac{2a}{1+a}$, obtained by choosing
three different glass transition points characterized by different
exponent parameters $\lambda$ introduced in Sect. 2.1. The exponents
$\tilde m$ vary by almost a factor of 3,
and the window in shear
rates, where Eq.~(\ref{neu4}) holds, shifts appreciably with
$\lambda$. The plot of $\dot \gamma^{1-\tilde m}
d\sigma/d\dot\gamma$ versus shear rate  shows an (almost)
constant plateau in the $\dot\gamma$-window, where Eq.~(\ref{neu4})
holds.

\section{Comparison with experiments}

\begin{figure}
\resizebox{1\columnwidth}{!}{\includegraphics{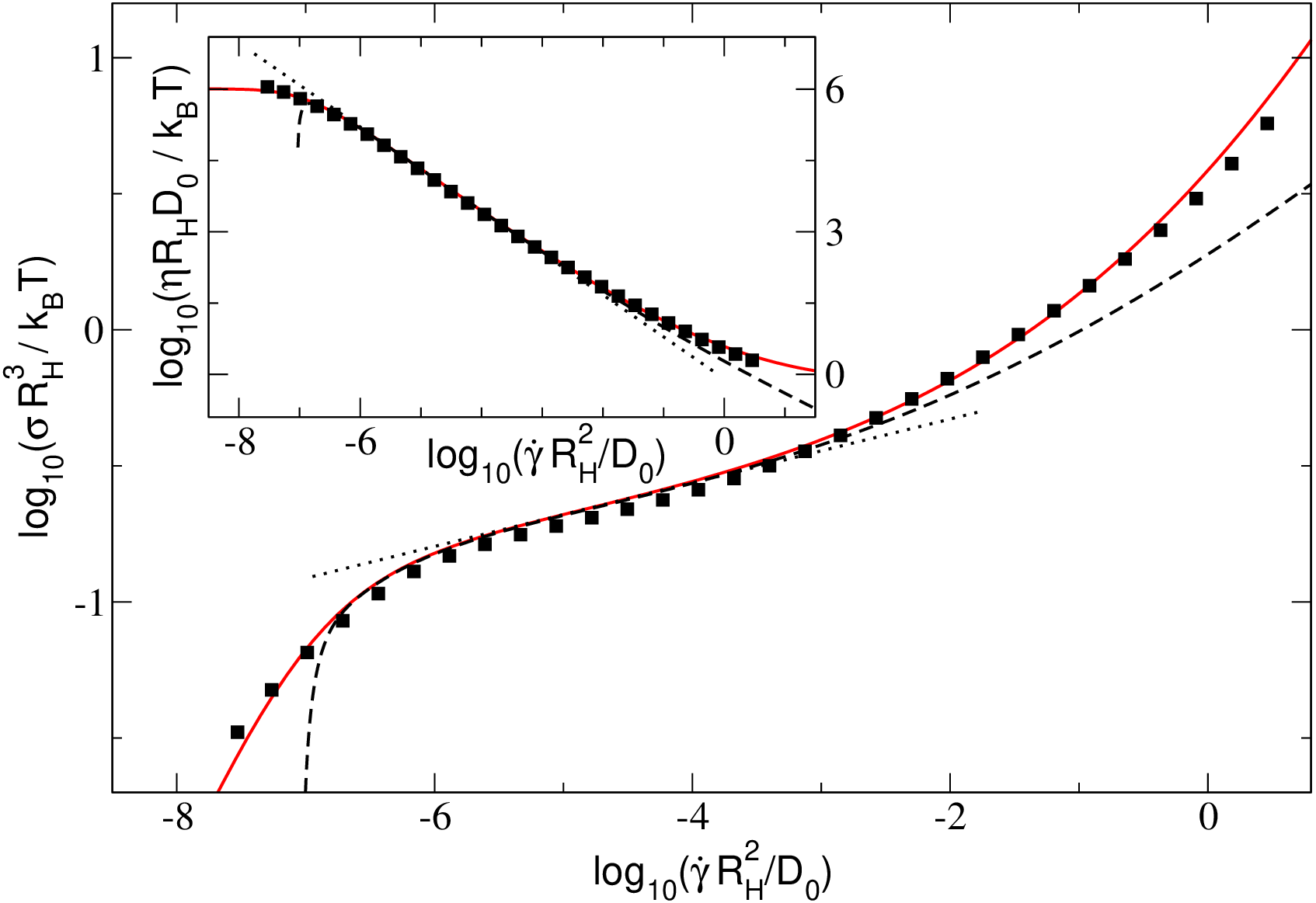}}
\caption{\label{fig:fig_exp1} Reduced flow curves for a core-shell
dispersion at an effective volume fraction of
$\phi_{eff}=0.580$. Here $R_H$ denotes the hydrodynamic radius and
$D_0$ the self diffusion coefficient of the colloidal particles;
$k_BT $ is the thermal energy. The solid line (red) shows the result
for the fitted $F_{12}^{\left(\dot{\gamma}\right)}$-model with
$v_2^c=2.0$. The fitted parameters are: $\varepsilon=-0.00042$,
$\gamma_c=0.14$, $v_{\sigma}=70k_BT/R_H^3$, $\Gamma=80D_0/R_H^2$ and
$\eta_{\infty}=0.394k_BT/R_HD_0$. The dashed line shows the
corresponding result for the $\Lambda$-formula. The dotted line
shows the inflection tangent of the numerically determined flow
curve with a slope of $p=0.12$.
The inset shows the corresponding results for the viscosity.
}
\end{figure}

\begin{figure}
\resizebox{1\columnwidth}{!}{\includegraphics{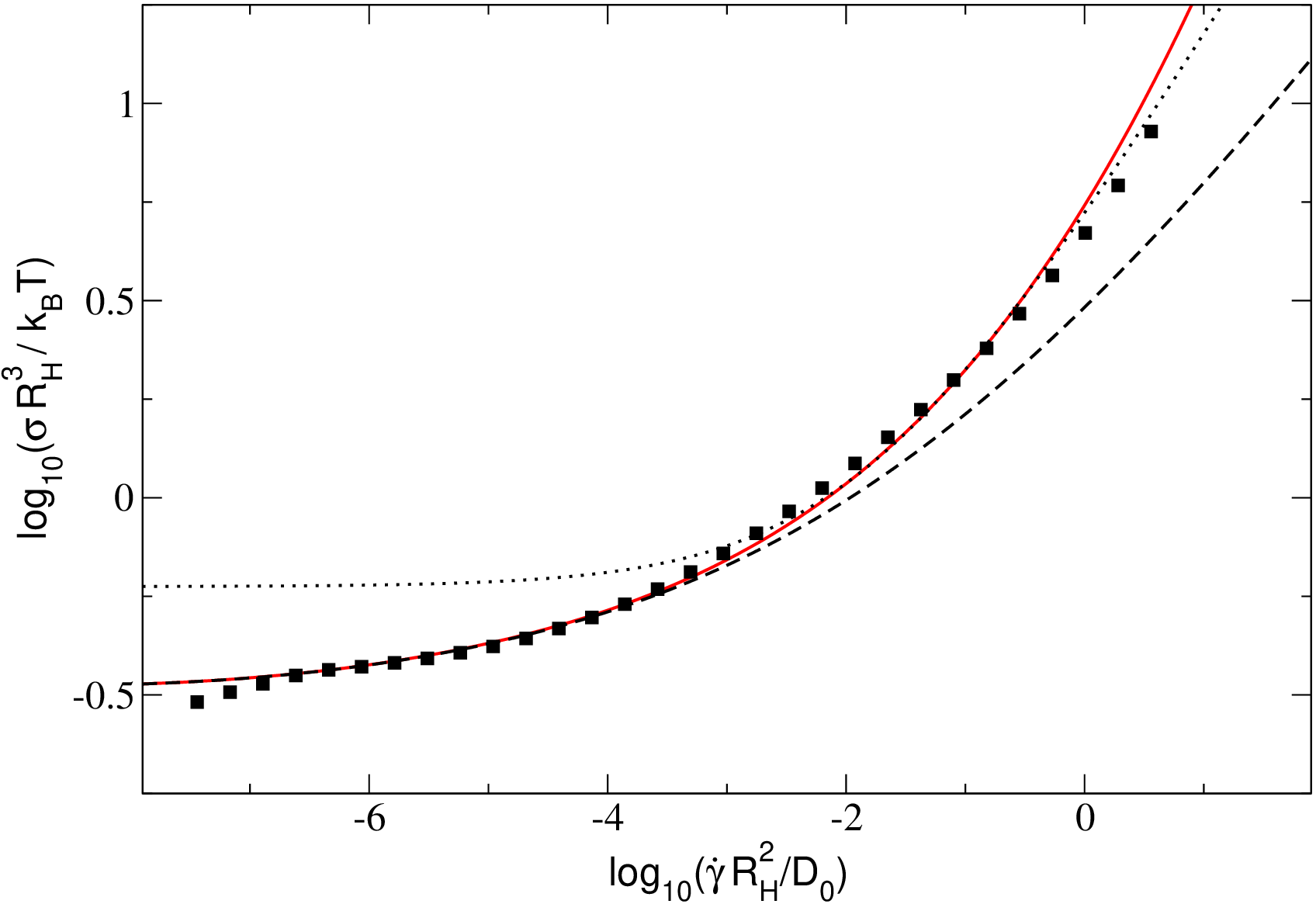}}
\caption{\label{fig:fig_exp2} Reduced flow curves for a core-shell
dispersion at an effective volume fraction of
$\phi_{eff}=0.629$; quantities as defined in the caption of Fig.
\ref{fig:fig_exp1}. The solid line (red) shows the result for the
fitted $F_{12}^{\left(\dot{\gamma}\right)}$-model with $v_2^c=2.0$.
The fitted parameters are: $\varepsilon=0.000021$, $\gamma_c=0.16$,
$v_{\sigma}=115k_BT/R_H^3$, $\Gamma=120D_0/R_H^2$ and
$\eta_{\infty}=0.431k_BT/R_HD_0$. The dashed line shows the
corresponding result for the $\Lambda$-formula. The dotted line
shows the fitted Herschel-Bulkley law given by Eq. (\ref{neu4}) with
the analytically calculated exponent $\tilde m=0.489$. }
\end{figure}

Fig. \ref{fig:fig_exp1} and Fig. \ref{fig:fig_exp2}  show
experimental data recently obtained by Siebenb\"urger et al.
\cite{Sie:08} on polydisperse dispersions of thermosensitive
core-shell particles \cite{Cra:06}. In all cases stationary
states were achieved after shearing long enough, proving that ageing
could be neglected even for glassy states. Along the lines of the
work \cite{Cra:07}, the $F_{12}^{\left(\dot{\gamma}\right)}$-model
was used to fit both the flow curves and the linear viscoelastic
moduli simultaneously, which strongly restricted the choice of  fit
parameters. Here we only show two representative results for the
flow curves, comparing the asymptotic results.

Fig. \ref{fig:fig_exp1} shows the result for a liquid-like flow
curve where the $\Lambda$-formula holds for approximately four
decades. The pseudo power law resulting from the inflection tangent
of the flow curve holds for approximately two decades within the
range of validity of the $\Lambda$-formula.

Fig. \ref{fig:fig_exp2} shows the result for a flow curve, where a
small positive separation parameter was necessary to fit the flow
curve and the linear viscoelastic moduli simultaneously. The
data fall below the fit curves for very small shear rates, which
indicates the existence of an additional decay mechanism neglected
in the present approach \cite{Cra:07,Sie:08}. Again, the
$\Lambda$-formula describes the experimental data correctly for
approximately four decades. For higher shear rates, the Herschel-Bulkley
law given by Eq. (\ref{neu4}) can be fitted in a window of
approximately two decades. We have derived Eq. (\ref{neu4}) for
$\varepsilon=0$, nevertheless the fit is possible for two reasons:
In the present case, the separation parameter is sufficiently small,
and for higher shear rates the $\varepsilon$-dependence of the flow
curves becomes negligible.

\section{Conclusion}

In the present work we have analyzed the shear flow behavior of
dense colloidal dispersions in the framework of the schematic
$F_{12}^{\left(\dot{\gamma}\right)}$-model based on the
mode-coupling theory. This model describes the transition from a
shear-thinning liquid to a yielding glass \cite{Fuc:03}. While in
the liquid state the steady state shear stress decays linearly to
zero for small shear rates, in the glassy state the shear stress
arrests on a finite dynamic yield stress value in the zero-shear limit.
In the liquid thus a linear response regime characterized by a
Newtonian viscosity exists. The viscosity diverges at the glass
transition according to Eq. (\ref{viscosity}) \cite{Goe:91}, and
carries a regular shear rate dependent correction of order
$\dot\gamma^2$. The glass, on the other hand, never exhibits a
regular linear response in shear rate. For high shear rates the
shear stress depends linearly on the shear rate for both cases.

Close to the glass transition and for small shear rates, we studied
the model using asymptotic expansions. The derivation of a quite
general asymptotic formula summarizing many aspects of the flow
curves, viz. the steady state shear stress as a function of the
shear rate, was the central result of our work. This formula
provides a deeper understanding of the shapes of the flow curves.
The scaling of the yield stress and its singular behavior at the
glass transition point can easily be obtained by studying special
asymptotic limits. The yield stress $\sigma^+$ jumps discontinuously
from zero in the fluid to a finite value at the glass transition,
and grows with the power law series from Eq.
(\ref{eq:yield_stress_2}) deeper into the glass. For all states
(deep) in the glass, the stress increases (non-analytically) with
the absolute value of the shear rate, see Eq.
(\ref{eq:sigma_linear}).  At the glass transition point our formula
reduces to the generalized Herschel-Bulkley law Eq.
(\ref{eq:crit_flow_2}) known from previous studies \cite{Fuc:03} and
\cite{Hen:05}. We have also demonstrated in Fig. \ref{fig:8} that
the apparent power law behavior of the flow curves, or of the
viscosity, in the liquid state for intermediate shear rates is only
an artifact of the inflection point occurring in a double
logarithmic plot. This inflection point necessarily occurs for any
model including shear thinning. The Herschel-Bulkley law of Eq.
(\ref{neu4}), which holds in a window of intermediate and not too
small shear rates, should be quite accessible experimentally.
Importantly, the exponent $\tilde m$ is connected to other exponents
characterizing the material properties, and can thus be inferred
from other measurements. Also, it can be calculated using Eq.
(\ref{eq:critical_exponent}) from first principles starting with the
particle interactions and using mode coupling theory \cite{Goe:91}.

In previous studies, the $F_{12}^{\left(\dot{\gamma}\right)}$-model
was successfully applied to describe experimental data for both the
steady state shear stress and the linear viscoelasticity, see
\cite{Cra:07} and \cite{Cra:06}. As shown in \cite{Hen:05} and
\cite{Var:06}, the model also describes simulation data for the
steady state shear stress with a high accuracy. Recently, the
Herschel-Bulkley law of Eq. (\ref{neu4}) with the predicted exponent
$\tilde m^{\rm HS}=0.473$ was observed in experiments on dispersions
of colloidal hard spheres \cite{monterey}. However, only the recent
experiments of Siebenb\"urger et al. \cite{Sie:08} provide  access
to shear rates and distances from the glass transition points which
are sufficiently small for a test of our asymptotic description of
the flow curves. Our asymptotic formulae describe these experiments
successfully in shear rate windows of several decades. Especially,
the pseudo power law is clearly observable in the experimental data.

\section{Acknowledgments}
We thank O. Henrich, J. J. Crassous, M. Siebenb\"urger, and M. Ballauff
for stimulating discussions. We especially thank M. Siebenb\"urger
and M. Ballauff for access to their experimental data prior to
publication. We acknowledge financial support by the Deutsche
Forschungsgemeinschaft in Transregio SFB Tr6 and IRTG 667.
\appendix

\section{Discussion of the $\Lambda$-function \label{sec:lambda}}

\begin{figure}
\resizebox{1\columnwidth}{!}{\includegraphics{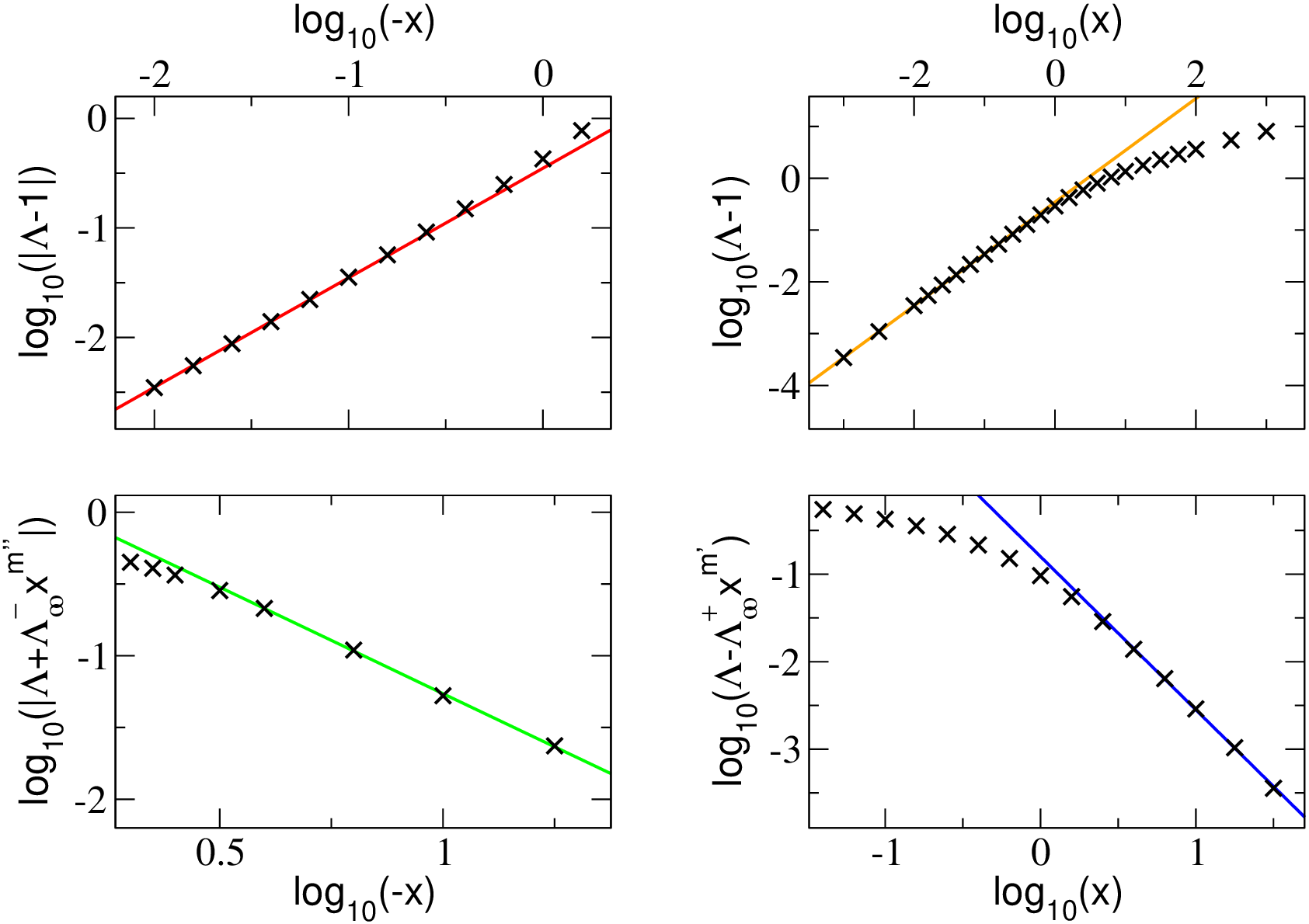}}
\caption{\label{fig:4} The next-to-leading order asymptotes of $\Lambda\left(x\right)$.
The crosses are numerically obtained data for $\Lambda\left(x\right)$.
The two upper panels show linear fits for small $\left|x\right|$.
The left lower panel shows the power law for $x\rightarrow-\infty$
with the fitted exponent $\xi$. The right lower panel shows for $x\rightarrow\infty$
the fit with the analytically calculated exponent $\zeta$.}
\end{figure}

In this section we present some numerical results
for the next-to-leading order asymptotes of $\Lambda\left(x\right)$.
For negative separation parameters, we have identified
\begin{equation}
\Lambda\left(x\rightarrow-\infty\right)=
-\Lambda_{\infty}^{-}\left|x\right|^{m''}+\Lambda_{1}^{\xi}\left|x\right|^{-\xi},
\label{eq:lambda_1}
\end{equation}
\begin{equation}
\Lambda\left(x\rightarrow0^{-}\right)=1+\Lambda_{1}^{-}x,
\label{eq:lambda_2}
\end{equation}
with an empirical exponent $\xi$. Inspired by this, we have found that $\Lambda\left(x\right)$
can be well approximated by using the ansatz
\begin{equation}
\Lambda\left(x\leq0\right)=\left\{ \begin{array}{rcl}
{\displaystyle \Lambda_{>}^{-}\left(x\right)} & , & {\displaystyle \left|x\right|>\chi}\\
\\{\displaystyle \Lambda_{<}^{-}\left(x\right)} & , & {\displaystyle 0\leq\left|x\right|\leq\chi}\end{array}\right.,
\label{eq:lambda_3}
\end{equation}
\begin{equation}
{\displaystyle \Lambda_{>}^{-}\left(x\right)}=-\Lambda_{\infty}^{-}\left|x\right|^{m''}+\sum_{n=1}^{N}\Lambda_{n}^{\xi}\left|x\right|^{-\xi n},
\label{eq:lambda_4}
\end{equation}
\begin{equation}
\Lambda_{<}^{-}\left(x\right)=1+\sum_{n=1}^{N}\Lambda_{n}^{-}x^{n}.
\label{eq:lambda_5}
\end{equation}
The constant $\chi$ is defined by $\Lambda\left(-\chi\right)=0$.
We obtain $\chi\approx2$. We choose $N=3$ and determine $\Lambda_{n}^{\xi}$
and $\Lambda_{n}^{-}$ by non-linear curve fittings. For positive
separation parameters, we have identified
\begin{equation}
\Lambda\left(x\rightarrow0^{+}\right)=1+\Lambda_{1}^{+}x,
\label{eq:lambda_6}
\end{equation}
\begin{equation}
\Lambda\left(x\rightarrow\infty\right)=\Lambda_{\infty}^{+}x^{m'}+\Lambda_{1}^{\zeta}x^{-\zeta}.
\label{eq:lambda_7}
\end{equation}
In addition, by using Eq. (\ref{eq:sigma_linear}) we have identified
\begin{equation}
\zeta=\frac{1-m}{\tilde{m}}\label{eq:zeta}.
\end{equation}
Inspired by these results, we have found that $\Lambda\left(x\right)$ can be well approximated
by
\begin{equation}
\Lambda\left(x\right)=\left\{ \begin{array}{rcl}
{\displaystyle \Lambda_{>}^{+}\left(x\right)} & , & {\displaystyle x>1}\\
\\{\displaystyle \Lambda_{<}^{+}\left(x\right)} & , & {\displaystyle 0\leq x\leq1}\end{array}\right.,
\label{eq:lambda_0}
\end{equation}
\begin{equation}
\Lambda_{>}^{+}\left(x\right)=\Lambda_{\infty}^{+}x^{m'}+\sum_{n=1}^{N}\Lambda_{n}^{\zeta}x^{-\zeta n},
\label{eq:lambda_9}
\end{equation}
\begin{equation}
\Lambda_{<}^{+}\left(x\right)=1+\sum_{n=1}^{N}\Lambda_{n}^{+}x^{n}.
\label{eq:lambda_10}
\end{equation}
Again, we choose $N=3$ and determine $\Lambda_{n}^{\zeta}$
and $\Lambda_{n}^{+}$ by non-linear curve fittings. The fact
that we obtain similar fitted values for $\Lambda_{n}^{+}$ and $\Lambda_{n}^{-}$
is an indication for a smooth behavior of $\Lambda\left(x\right)$
at $x=0$. With the presented formulae we are able to describe $\Lambda\left(x\right)$
for arbitrary arguments.
Fig. \ref{fig:3} and Fig. \ref{fig:4} summarize the results for
the $\Lambda$-function.

\section{Numerical parameters \label{sec:num}}

The numerical values for the parameters, which are relevant for our work, are summarized in the Tables \ref{tab:1}-\ref{tab:4}.

\begin{table}[ht]
\begin{centering}\begin{tabular}{|c|c||c|c||c|c||c|c|}
\hline
$v_{1}^{c}$&
$0.828$&
$\lambda$&
$0.707$&
$t_{0}$&
$0.426$&
$\hat{\tau}_{\eta}$&
0.240
\tabularnewline
\hline
$v_{2}^{c}$&
$2.00$&
$c^{\left(\dot{\gamma}\right)}$&
$0.586$&
$\hat{\tau}_{0}$&
$1.71$&
$G_{\infty}^c$&
$0.0858$
\tabularnewline
\hline
$f_{c}$&
$0.293$&
$\tilde{\dot{\gamma}}$&
$0.595$&
$\tilde{t}_0$&
$0.586$&
&
\tabularnewline
\hline
\end{tabular}\par\end{centering}
\caption{\label{tab:model} The parameters for the schematic $F_{12}^{\left(\dot{\gamma}\right)}$-model.
We have chosen $v_{2}^{c}=2.00$. The time scales $t_{0}$, $\hat{\tau}_{0}$ and $\hat{\tau}_{\eta}$
%and the elastic constant of the glass $G_{\infty}^c$
were calculated numerically. All other quantities were calculated
analytically.}
\label{tab:1}
\end{table}

\begin{table}[ht]
\begin{centering}\begin{tabular}{|c|c||c|c||c|c|}
\hline
$a$&
$0.324$&
$\tilde{m}$&
$0.489$&
$\zeta$&
$1.75$\tabularnewline
\hline
$b$&
$0.629$&
$m''$&
$2.32$&
$\xi$&
$1.48$\tabularnewline
\hline
$c$&
$0.586$&
$m'$&
$0.293$&
&
\tabularnewline
\hline
$\gamma$&
$2.34$&
$m$&
$0.143$&
&
\tabularnewline
\hline
&
&
$\bar{m}$&
$0.993$&
&
\tabularnewline
\hline
\end{tabular}\par\end{centering}
\caption{\label{tab:exponents} The numerical values for the exponents. The exponents $\zeta$ and $\xi$ were
first identified numerically. In addition, an analytical expression could be verified numerically for $\zeta$ which was
used to calculate it. All other
exponents could be derived analytically. Hence we can state that these exponents and $\zeta$
are only dependent on the exponent parameter $\lambda$, in this sense they are universal. This statement could not be clearly verified for $\xi$.}
\label{tab:2}
\end{table}

\begin{table}[ht]
\begin{centering}\begin{tabular}{|c|c||c|c||c|c|}
\hline
$\Lambda_{\infty}^{-}$&
$0.104$&
$\Lambda_{\infty}^{+}$&
$1.20$&
$\Lambda_{1}^{-}$&
$0.348$
\tabularnewline
\hline
$\Lambda_{1}^{\xi}$&
$1.81$&
$\Lambda_{1}^{\zeta}$&
$0.165$&
$\Lambda_{1}^{+}$&
$0.347$
\tabularnewline
\hline
$\Lambda_{2}^{\xi}$&
$-1.21$&
$\Lambda_{2}^{\zeta}$&
$-0.110$&
$\Lambda_{2}^{-}$&
$-0.0649$
\tabularnewline
\hline
$\Lambda_{3}^{\xi}$&
$-1.00$&
$\Lambda_{3}^{\zeta}$&
$0.0418$&
$\Lambda_{2}^{+}$&
$-0.0653$
\tabularnewline
\hline
&
&
&
&
$\Lambda_{3}^{-}$&
$0.0146$
\tabularnewline
\hline
&
&
&
&
$\Lambda_{3}^{+}$&
$0.00964$
\tabularnewline
\hline
\end{tabular}\par\end{centering}
\caption{\label{tab:Lambda} Numerically determined amplitudes for the leading
asymptotes and the empirical correction terms for the $\Lambda$-function.
The fact that we obtain similar fitted values for $\Lambda_{n}^{-}$
and $\Lambda_{n}^{+}$ is an indication for a smooth behavior of $\Lambda\left(x\right)$
at $x=0$.}
\label{tab:3}
\end{table}

\begin{table}[ht]
\begin{centering}\begin{tabular}{|c|c||c|c||c|c|}
\hline
$\sigma_{0}^{+}$&
$0.0148$&
$\sigma_{0}^{0}$&
$0.0148$&
$\sigma_{0}^{-}$&
$0.0148$\tabularnewline
\hline
$\sigma_{1}^{+}$&
$0.0591$&
$\sigma_{1}^{0}$&
$0.0495$&
$\sigma_{1}^{-}$&
$0.00517$\tabularnewline
\hline
$\sigma_{2}^{+}$&
$0.222$&
$\sigma_{2}^{0}$&
$0.155$&
$\sigma_{2}^{-}$&
$0.00169$\tabularnewline
\hline
$\sigma_{3}^{+}$&
$0.598$&
$\sigma_{3}^{0}$&
$0.350$&
$\sigma_{3}^{-}$&
$0.000399$\tabularnewline
\hline
$c_{2}^{+}$&
$0.941$&
$c_{2}$&
$0.936$&
&
\tabularnewline
\hline
$c_{3}^{+}$&
$0.635$&
$c_{3}$&
$0.632$&
&
\tabularnewline
\hline
$\varepsilon_{*}$&
$0.00887$&
$\dot{\gamma}_{*}$&
$0.000506$&
&
\tabularnewline
\hline
\end{tabular}\par\end{centering}
\caption{\label{tab:flow_curves} Numerically determined amplitudes for the
flow curves. The only fitted parameters are $\sigma_{n}^{+}$ from
which the other quantities were calculated.}
\label{tab:4}
\end{table}

\end{document}